\documentclass[manuscript]{aastex62}
\usepackage{graphicx}
\usepackage[caption=false]{subfig}
\usepackage{amsmath}
\usepackage{amssymb}
\usepackage{xcolor}

\newcommand{\specialcell}[2][c]{%
	\begin{tabular}[#1]{@{}c@{}}#2\end{tabular}}

\received{Oct 28, 2018}
\accepted{Aug 1, 2019}

\submitjournal{ApJ}

\shorttitle{HII Regions and SCUBA-2 Cores}
\shortauthors{Bobotsis and Fich }

\begin{document}

\title{The Distribution of Dense Cores near HII Regions}

\author{George Bobotsis}
\affiliation{Department of Physics \& Astronomy \\
University of Waterloo \\
200 University Ave. West  \\
Waterloo, Ontario, Canada}

\author{Michel Fich}
\affiliation{Department of Physics \& Astronomy \\
University of Waterloo \\
200 University Ave. West  \\
Waterloo, Ontario, Canada}



\begin{abstract}
An investigation of dust emission associated with a large sample of HII regions has been carried out. Stacked results from this sample suggest that each HII region is at or near the center of a cluster of dense cores, that extends far beyond the HII region, and has volume density that decreases as $r^{-3}$. The data also shows evidence for enhanced numbers of cores near the boundary of the HII regions. At the same time, a significant decrease in the number of cores, consistent with no cores, is observed in the interior of these HII regions. Neither these HII regions, nor their associated massive OB stars were found to have a significant heating effect on their associated dusty clumps. \textquotedblleft Clouds", or the outermost layers of the clumps in which the cores are embedded, are found to exert a strong shielding effect to external heating sources. Despite this, a large portion of the identified cores was found to be warmer than their surrounding cloud and consequently may be in the initial stages of star formation. The star formation efficiency of the 7 HII region systems with the most reliable mass budgets ranged between 1\% and 9\%.
\end{abstract}



\section{Introduction} \label{sec:intro}
Much of the study of the formation of stars focusses on the process of how a denser part of the \textquotedblleft Interstellar Medium" (ISM) evolves into a pre-main sequence star. However few stars form in isolation. There have been a number of studies in the past investigating the feedback of individual HII regions including some that find evidence for feedback and other studies that have found no or very limited signs of feedback. In this paper, we use submillimeter wavelength observations of dust emission to look at the relationship between HII regions and the dense interstellar material associated with them. We measure the physical properties of this material, determine how it is distributed around the HII regions, and look for signs of interactions taking place. Overall, we were able to amass a large number of observations of many HII regions in order to set some constraints on their feedback in general as well as to understand how important feedback is on average.
\\ \\
An HII region is the product of one or more massive, OB stars embedded in a molecular cloud. The ionizing radiation provided by the parent OB star(s) not only drives the outward expansion of the HII region, but can also heat nearby clumps while progressively eroding them away by gradually ionizing them. Sandford et al. (1982) have suggested that dusty clumps bombarded by ionizing radiation can be lead to collapse due to the large radiation pressure exerted on their outer layers, but also, the large extinction of the dusty content delays the ionization of gas in their interior regions. These claims have been followed up with analytic models (Kovalenko \& Shchekinov 1992) as well as simulations (Motoyama, Umemoto, \& Shang 2007), (Bisbas et al. 2011). This process of forming stars is now commonly referred to as \textquotedblleft Radiative Driven Implosion" (RDI).
\\ \\
Furthermore, older (post-Str\"{o}mgren-sphere) HII regions have slowly-expanding, shockwave-ionization-front structures, whose propagation can have dramatic effects on the local ISM. More specifically, the pressure differential between the shockwave-ionization-front bundle and the local ISM is large and sharp, leading material encountered along the way to become swept-up and compressed into clumps. Elmegreen and Lada (1977) suggested that these clump condensations can become gravitationally unstable as they grow beyond a certain column density, and collapse to form stars. This process of forming stars is now commonly referred to as Collect and Collapse (CC). In later work, Lada used this process to explain how star formation could propagate in sequential waves through a \textquotedblleft Giant Molecular Cloud" (GMC) where essentially swept up molecular material forms a shell filament along the boundary of an expanding HII region. Gravitational instabilities can develop along this filament, leading to its fragmentation into several dense cores, which ultimately collapse to give rise to massive stars. These maintain the velocity of the original layer they formed in, and give rise to subsequent HII regions deeper within the cloud structure. In later work, Lada (1987) describes the end to this chain as the point at which the massive stars completely strip their parent GMC of gas and dust material.
\\ \\
One of the first HII regions observed to have extended molecular condensations forming along its boundary was Sh-2 104, where emission from various molecular CO transitions was used to trace molecular material while radio-continuum emission was used to trace the location of the HII region itself (Deharveng et al. 2003). With the rise of submillimeter astronomy, an increasing number of similar HII regions were found using the cooler dust components to trace the associated gas. Some of these include the case of RCW 79 (Zavagno et al. 2005), Sh-2 219 (Deharveng et al. 2006), RCW 120 (Zavagno et al. 2007), Sh-2 212 (Deharveng et al. 2008), the Sh-2 254 -- Sh-2 258 complex (Chavarria et al. 2008), Sh-2 217 (Brand et al. 2011), Sh-2 90 (Samal et al. 2014), Sh-2 39 (Duronea et al. 2017) and Sh-2 242 (Dewangan et al. 2017), all of which show signs of feedback acting on material around the HII region. On the other hand, in a study of 25 HII regions conducted by Xu et al. (2014) only 3 had timescale estimates that allowed for the action of feedback.
\\ \\
In this work we assemble observations of dust emission from 38 SCUBA-2 images containing a total of 53 HII regions. Many of these results are only accessible through the use of such a large sample, which, to the best of our knowledge, has never been available before. The HII regions comprising this sample are all larger, more evolved, mature objects, in which the effects of feedback are expected to be most pronounced.

\section{Observations}
The sample of HII region systems analyzed is comprised exclusively of mature, galactic HII regions taken from the \textquotedblleft Sharpless" (Sh-2) (Sharpless 1959) and \textquotedblleft Blitz-Fich-Stark" (BFS) (Blitz, Fich, \& Stark 1982) catalogues. These catalogues select HII regions from the Palomar Sky Survey, a set of optical large (6 degree) images of the sky. This results in a sample of objects that have evolved beyond the compact stage where they are still deeply embedded in, and obscured by,  dense clouds. The smallest HII regions visible on these images are approximately half an arcminute in angular size but the largest may cover several degrees. Our sample is drawn from those that have VLA observations available. This in practice limits the sample to objects less than $\approx20'$ in diameter.
\\ \\
The properties of these HII regions are determined using 1.46 GHz and 4.89 GHz VLA data (Fich 1986, 1993). This data was only available for 48 of the 53 HII regions located on these SCUBA-2 images. The position and size of the HII regions is determined using the 10\% flux contour of their radio-continuum emission, which is subsequently fitted to a circle. Number densities are determined using the simplified expression from Mezger \& Henderson (1967) along with small corrections for the radio observing band and the use of a cylindrical approximation, and an electron temperature of $\approx 8 \times 10^3 \ K$, equal to the average electron temperatures of all HII regions analyzed in the work of Rudolph et al. (2006). To determine the physical radii used in this expression, radial distances provided by Foster \& Brunt (2015) are used, and when those aren't available, older estimates compiled by Chan \& Fich (1995) are used instead. The masses are determined by incorporating the overall number density, and assuming an abundance ratio $He^{++}/H^{+} \approx 0$ and $He^{+}/H^{+} = He/H \approx 0.06$ (Rudolph et al. 2006).
\\ \\
For the identification of submillimeter-emitting clumps in the vicinity of these HII regions, 450$\mu$m and 850$\mu$m SCUBA-2 data was used, which was collected from various projects from the \textquotedblleft Canadian Astronomical Data Center" (CADC) archives and reduced using a standard data reduction pipeline procedure (Holly \& Currie 2014). The resolution of the SCUBA-2 instrument allowed the separation of these clumps into a warm outer layer \textquotedblleft cloud" and one or more inner, dense condensations \textquotedblleft cores".
\\ \\
Photometry of the clumps was obtained using a scripted routine in Python. The routine treats the cloud and embedded cores individually, with the cloud flux always removed from that of the embedded cores. In addition, negative bowl artifacts persisting after the application of a mask during data reduction are treated on a clump-by-clump basis. This is done by approximating the residual negative bowl as a step function, whose value is estimated using small, circular apertures that are employed along the periphery of the source with the goal of finding the most negative mean flux per pixel value occurring there. Once found, this value is used to characterize the negative bowl, and allows its removal from the source flux. Finally, the contamination in the 850$\mu$m band from the molecular CO(3-2) transition is treated using a correction factor of 10\%. The choice for this value comes from consideration of SCUBA-2 surveys of NGC 1333, NGC 2071 and NGC 2024 by Drabek et al. (2012) in which the majority of SCUBA-2 sources experienced contamination levels less than 20\%, but also, a survey of the Taurus star-forming region by Buckle et al. (2015) within which all SCUBA-2 sources experienced a contamination level less than 15\%, with a large number of sources not exceeding a contamination level of 5\%.
\\ \\
\begin{figure}
	\centering 
	\subfloat[Sh-2 168]{
		\fbox{\includegraphics[width=0.48\columnwidth]{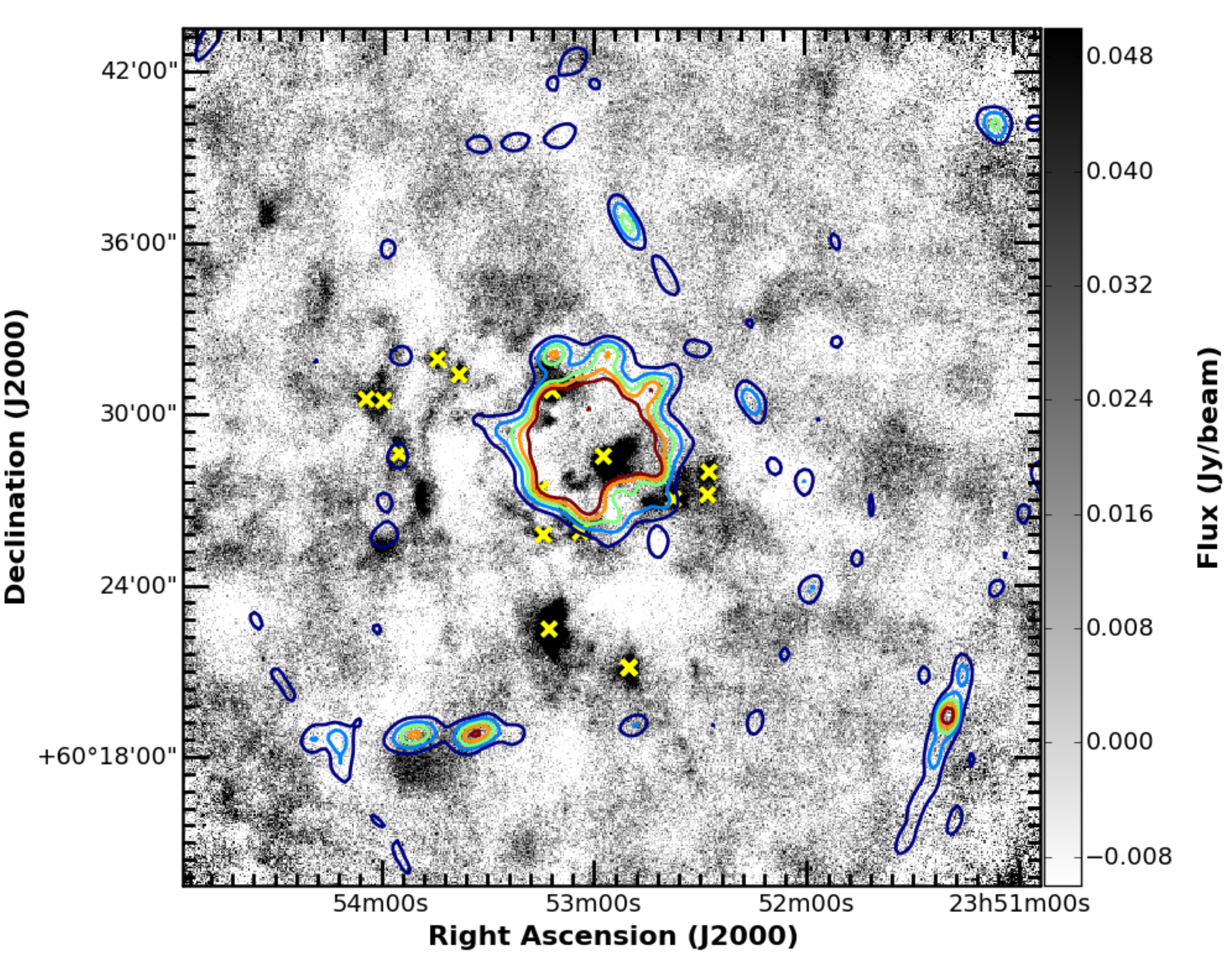}}}
	\subfloat[Sh-2 201]{
		\fbox{\includegraphics[width=0.48\columnwidth]{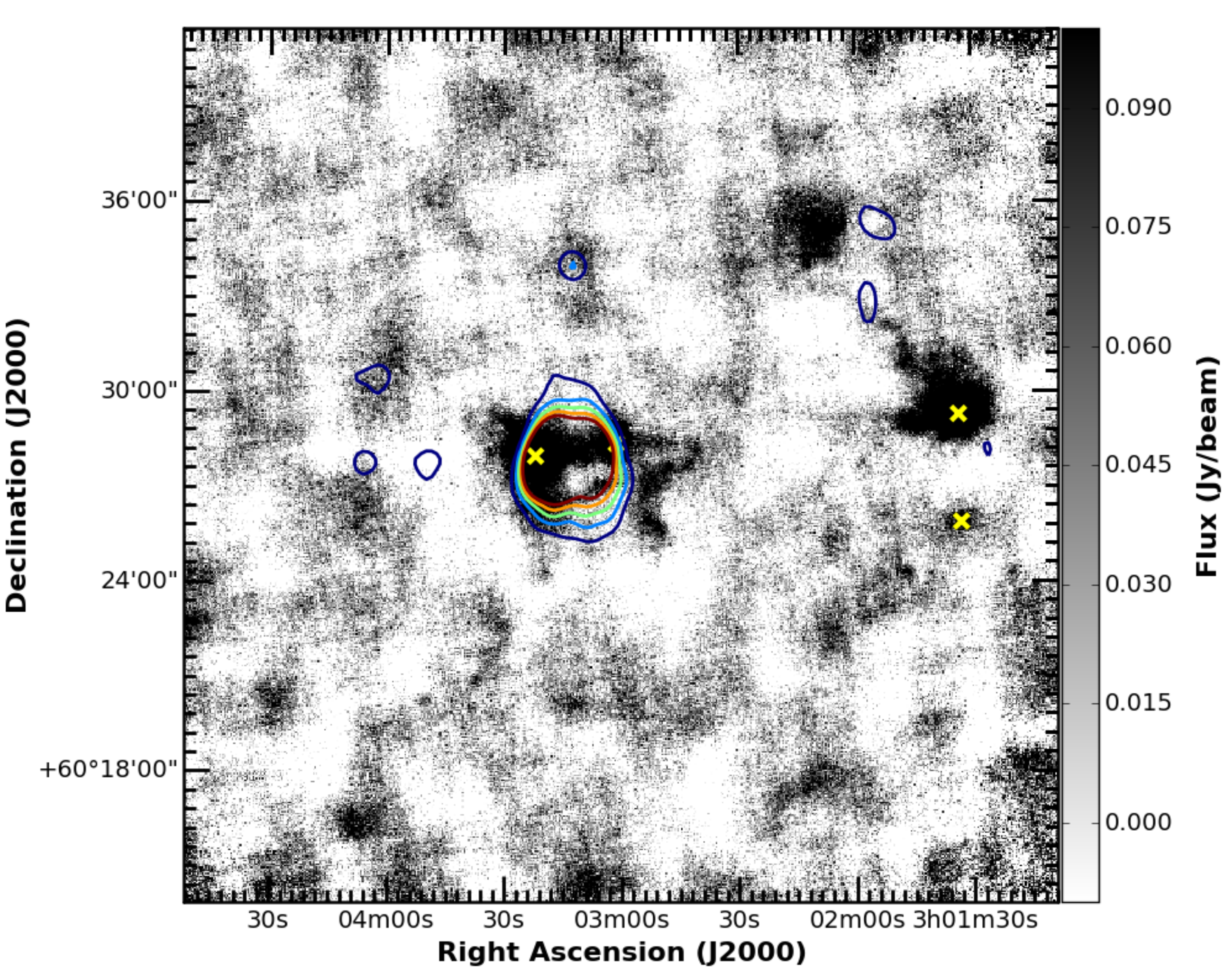}}}
	
	\subfloat[Sh-2 242]{
		\fbox{\includegraphics[width=0.48\columnwidth]{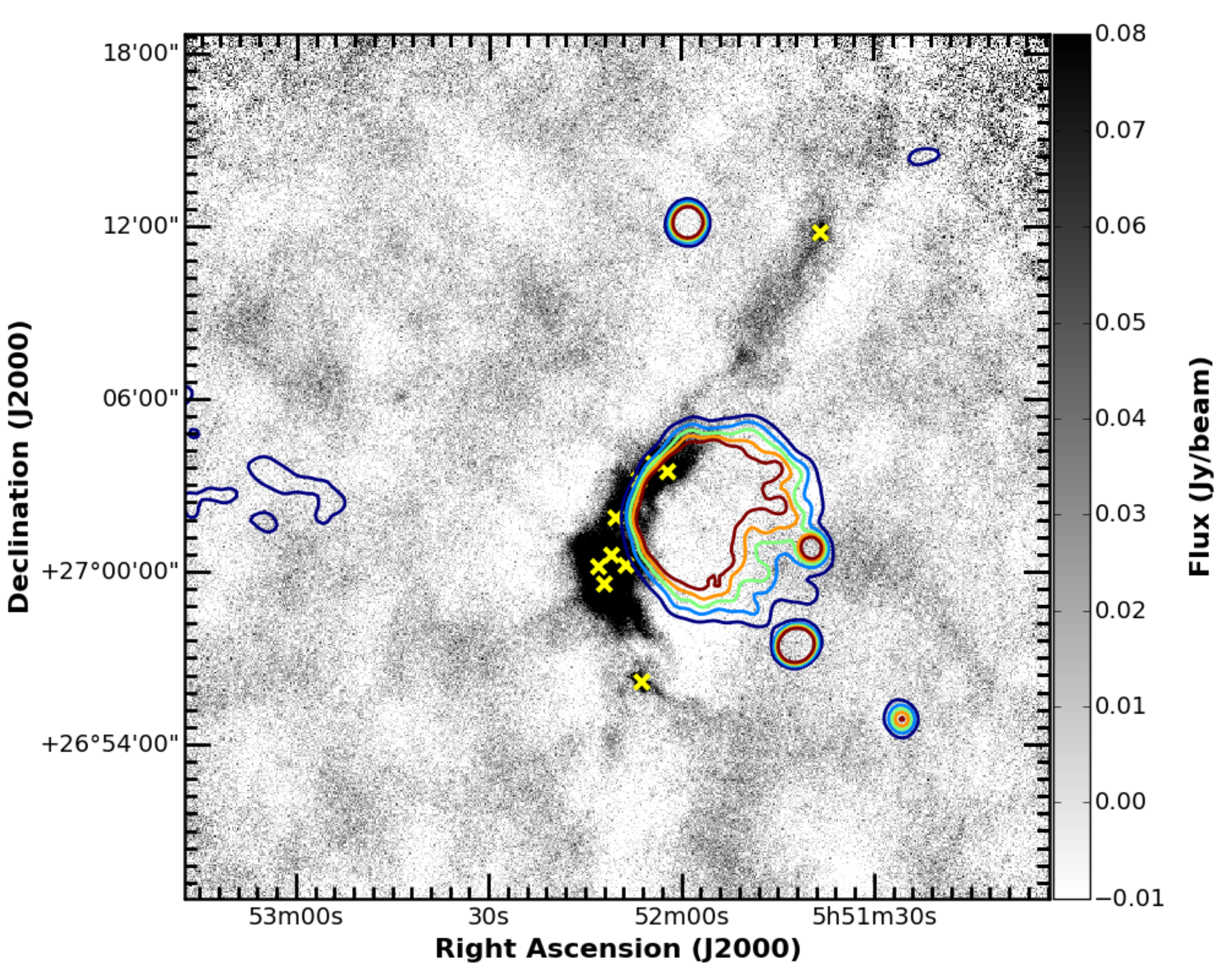}}}
	\subfloat[Sh-2 305]{
		\fbox{\includegraphics[width=0.48\columnwidth]{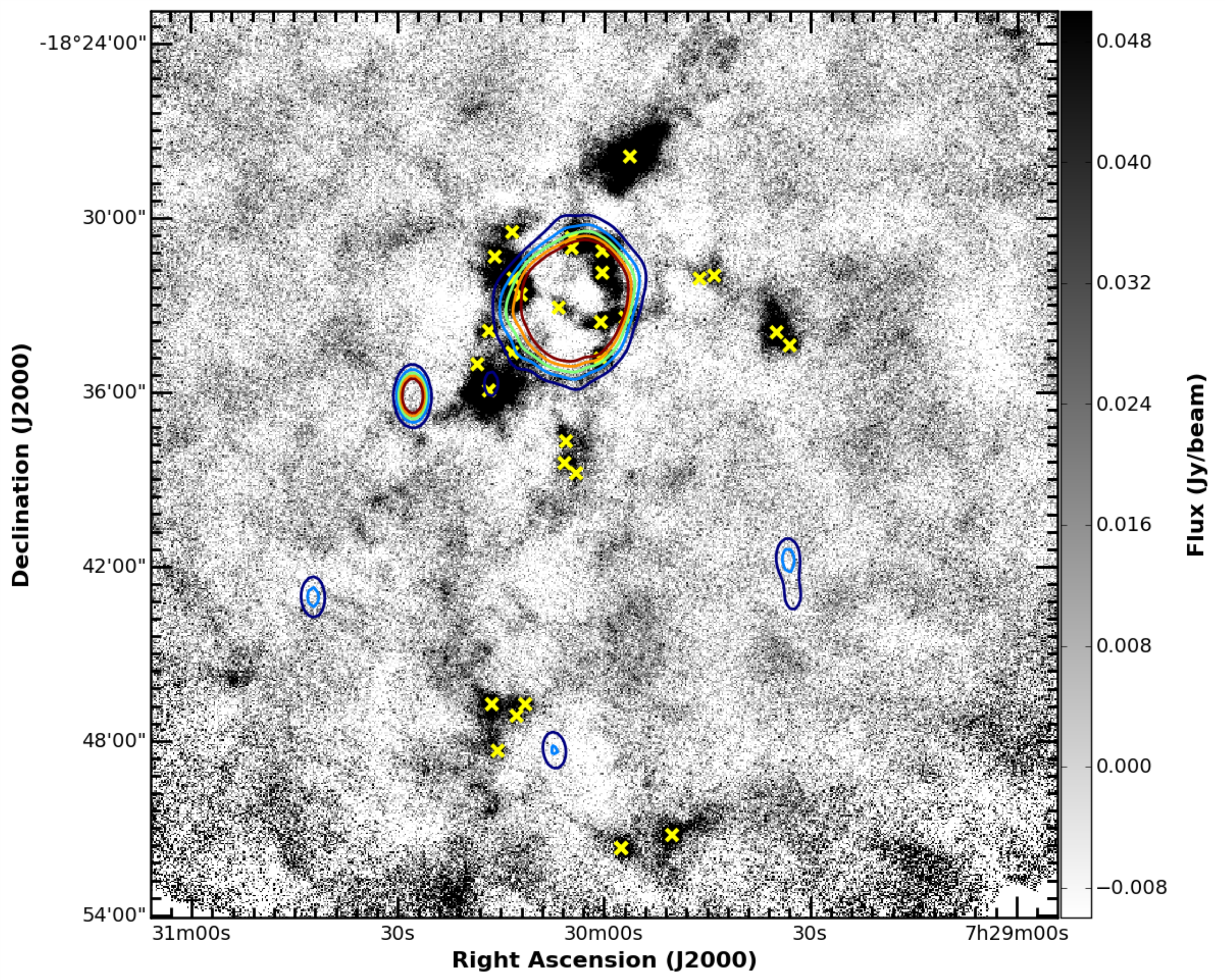}}}
	\caption{A collection of 4 HII regions representing the variety of SCUBA-2 condensation morphologies encountered in this work. The images consist of SCUBA-2 850$\mu$m emission overlaid with VLA 1.46 GHz contours. The contours are color-coded based on confidence level multiples (blue/cyan/green/orange/red $\rightarrow$ 1$\sigma$,2$\sigma$, 3$\sigma$, 4$\sigma$, $5\sigma$). The yellow "x" ticks indicate the location of identified cores.}
	\label{fig:HII_regions}
\end{figure}
The opacity model used is the one originally proposed by Ossenkopf \& Henning (1994) for use with protostellar cores. Due to the limitations from the consideration of only two submillimeter bands (450$\mu$m and 850$\mu$m), a prescribed value of $\beta = 1.8$ is used along with this opacity model. A dust to gas mass ratio of 1:100 and a composition of 70\% $H_{2}$ by-mass are also assumed throughout these calculations.
\\ \\
A comprehensive list of properties was made for all identified cores and clouds of this sample. This included measured properties such as positions, size and 450$\mu$m/850$\mu$m flux, but also several derived properties, such as average temperature, total mass, average column density and average number density using the same recipes as those used for the analysis of SCUBA-2 clumps in surveys of the Gould Belt (Buckle et al. 2015).
\\ \\
From the total of 38 SCUBA-2 450$\mu$m and 850$\mu$m systems investigated, 31 (82\%) had one or more dusty clumps identified in the vicinity of an HII region from the considered sample. A total of 185 clumps and 333 cores were identified, and after discarding a few of these objects on the premise that they were too far from the nearest HII region for an association to exist, 176 clumps (95\%) and 315 embedded cores (95\%) continued to analysis.
\\ \\
We portray the variety of structures we encountered with a set of 4 representative cases in Figure \ref{fig:HII_regions}.  The most commonly seen structure (58\% of our fields) have 4 or fewer dense cores embedded in a few clumps as seen in the object Sh-2 201 in this Figure. 81\% of our image fields have 20 or fewer dense cores, with a few cases where most or all of the cores are embedded in one massive clump as in Sh-2 242 in Figure \ref{fig:HII_regions}. Only six of our fields (the remaining 19\%) show large numbers of cores (28 to 49 cores) as seen in Sh-2 168 and Sh-2 305 in Figure \ref{fig:HII_regions}.  However these six fields contain 69\% of the dense cores in our sample. Many of the dense cores are found far beyond the boundary of the HII region: we examine the radial distribution of the positions of these cores below.
\section{Results}
\subsection{Summarized Properties}
Our sample was selected to include virtually all of the HII regions with data in the SCUBA-2 archives. Nonetheless, we excluded a handful of more extreme HII regions, such as very nearby and/or large in angular extent (e.g. the Orion HII region). Also, our sample was not selected by prioritizing similarity, such that it covered a small range in distance, angular size or brightness. Despite this, the measured and calculated properties for this sample were remarkably uniform. Table \ref{table:Results} summarizes the properties of the sample, including median values, the \textquotedblleft central population" (i.e the range of each property when outliers are not considered) and finally, the full range.
\begin{table}[!htb]
	\centering

	\caption{Table of summarized properties for all HII regions and their associated clouds and cores.}
	
	\begin{footnotesize}
	\setlength\tabcolsep{2pt}

		\begin{tabular}{|l|c|c|c|c|}	
		\hline
		\textbf{Property} & \textbf{Median} & \textbf{Range (Excluding Outliers)} & \textbf{Sample Fraction} & \textbf{Full Range} \\
		\hline
		Distance (kpc) & $3.9$ & $2 \leq d \leq 9$ & $87\%$ & $1.15 \leq d \leq 12.6$ \\
		\hline
		HII Region Angular Radius ($''$) & $100$ & $30 \leq R \leq 300$ & $90\%$ & $24 \leq R \ \leq 600$ \\
		\hline
		HII Region Physical Radius (pc) & $2.20$ & $0.5 \leq R \leq 5$ & $80\%$ & $0.35 \leq R \leq 20.9$ \\
		\hline
		HII Region $n_{e}$ ($cm^{-3}$) & $40.7$ & $10 \leq n_{e} \leq 100$ & $79\%$ & $7.62 \leq n_{e} \leq 574$ \\
		\hline
		HII Region Total Mass ($M_{\odot}$) & $36.6$ & $2 \leq M \leq 300$ & $84\%$ & $0.32 \leq M \leq 8500$ \\
		\hline
		Cloud Angular Radius ($''$) & $36$ & $20 \leq R \leq 90$ & $96\%$ & $18 \leq R \leq 186$ \\
		\hline
		Cloud Physical Radius (pc) & $0.64$ & $0.2 \leq R \leq 2$ & $95\%$ & $0.17 \leq R \leq 7.6$ \\
		\hline
		Cloud 450$\mu$m Integrated Flux (Jy) & $7.27$ & $1 \leq F_{450} \leq 40$ & $83\%$ & $0.314 \leq F_{450} \leq 421$ \\
		\hline
		Cloud 850$\mu$m Integrated Flux (Jy) & $0.826$ & $0.2 \leq F_{850} \leq 5$ & $80\%$ & $0.059 \leq F_{850} \leq 49$ \\
		\hline
		Cloud Average Temperature (K) & $15$ & $9 \leq T \leq 30$ & $82\%$ & $7.4 \leq T \leq 243$ \\
		\hline
		Cloud Total Mass ($M_{\odot}$) & $106.5$ & $10 \leq M \leq 2000$ & $81\%$ & $2.21 \leq M \leq 15700$ \\
		\hline
		Cloud Average $N_{H_{2}}$ ($10^{21} cm^{-2}$) & $2.36$ & $0.5 \leq N_{H_2} \leq 6$ & $81\%$ & $0.045 \leq N_{H_2} \leq 32$ \\
		\hline
		Cloud Average $n_{H_{2}}$ ($cm^{-3}$) & $727$ & $150 \leq n_{H_{2}} \leq 2000$ & $80\%$ & $8.31 \leq n_{H_{2}} \leq 15300$ \\
		\hline
		Core Angular Radius ($''$) & $12$ & $10 \leq R \leq 20$ & $85\%$ & $4 \leq R \leq 54$ \\
		\hline
		Core Physical Radius (pc) & $0.26$ & $0.1 \leq R \leq 0.5$ & $86\%$ & $0.04 \leq R \leq 2.2$ \\
		\hline
		Core 450$\mu$m Integrated Flux (Jy) & $1.41$ & $0.2 \leq F_{450} \leq 7$ & $82\%$ & $0.08 \leq F_{450} \leq 307$ \\
		\hline
		Core 850$\mu$m Integrated Flux (Jy) & $0.196$ & $0.04 \leq F_{850} \leq 0.8$ & $81\%$ & $0.023 \leq F_{850} \leq 41.4$ \\
		\hline
		Core Average Temperature (K) & $19.4$ & $10 \leq T \leq 40$ & $83\%$ & $6.6 \leq T \leq 219$ \\
		\hline
		Core Total Mass ($M_{\odot}$) & $22.7$ & $2 \leq M \leq 150$ & $84\%$ & $0.37 \leq M \leq 12300$ \\
		\hline
		Core Average $N_{H_{2}}$ ($10^{21}cm^{-2}$) & $3.7$ & $1.5 \leq N_{H_{2}} \leq 15$ & $80\%$ & $0.253 \leq N_{H_{2}} \leq 35$ \\
		\hline
		Core Average $n_{H_{2}}$ ($cm^{-3}$) & $2990$ & $1000 \leq n_{H_{2}} \leq 15000$ & $80\%$ & $247 \leq n_{H_{2}} \leq 197000$ \\
		\hline
		\label{table:Results}
		\end{tabular}
	\end{footnotesize}
\end{table}
\\ \\
The measured quantities in Table \ref{table:Results} (angular sizes, fluxes) have uncertainties that are typically $\leq$ 10$\%$. Properties that depend on distance (physical radii, HII region density and mass) typically are uncertain by twenty percent. Values that depend on the ratio of the submillimeter fluxes (temperature, mass, and densities) are uncertain by 20-30$\%$ for the lower temperature objects (e.g. for T$\le$15K) but can be uncertain by 100$\%$ on the upper bound side due to the strong, non-linearity of the temperature-flux-ratio relationship at higher temperatures.
\\ \\
None of these quantities were normally distributed: there was always a strong skew to their distributions, usually with several extreme outliers. However, after removing the small number of such outliers these properties were found to have a relatively small range where most (typically over 80\%) of the sample was found to form a central population. For example, their distances from the Sun varied by more than a factor 10 and the angular radii of the HII regions varied by a factor of 25. Nonetheless, the physical radii of 80\% of these were within a factor of 10 of each other (and within a factor of 4 of the median).
\\ \\
The central population of most calculated properties varied by roughly one order of magnitude. The noticeable exceptions to this rule are the total mass and the average number density of the clouds and cores. This is likely because the calculation of these is very sensitive to the precision of the distance provided, the value of which is prone to larger uncertainties.
\\ \\
The central population of the HII region masses also varied by more than one order of magnitude. This is mostly due to the large span of sizes in the HII region sample considered. Note that the mass of the HII regions is used later in this paper in order to calculate star formation efficiency. However, the gas mass contributed by the HII regions is most commonly far less than that from other gaseous components (i.e the clouds and cores).
\\ \\
From the 53 HII regions analyzed, all had a distance, an angular radius and a physical radius estimate; while 48 (91\%) had an electron number density and total mass estimate. Furthermore, from the total of 176 clouds and 315 cores analyzed, all had an angular and physical radius estimate; all clouds and 275 (87\%) cores had a 450$\mu$m integrated flux measurement; all clouds and 312 (99\%) cores had an 850$\mu$m integrated flux measurement; 136 (77\%) clouds and 206 (65\%) cores had an average temperature estimate; 129 (73\%) clouds and 192 (61\%) cores had a total mass estimate; 133 (76\%) clouds and 203 cores (64\%) had an average $H_{2}$ column density estimate; and finally 134 (76\%) clouds and 203 (64\%) cores had an average $H_{2}$ number density estimate.
\\ \\
A detailed listing of all HII regions and associated clouds and cores along with their individual properties and their accompanying SCUBA-2 450$\mu$m and 850$\mu$m images can be found in the unpublished MSc thesis of Bobotsis (2018). A detailed discussion of all sources of uncertainty can also be found there.

\subsection{Core Number Counts} \label{sec:enhanced_condensation}
\begin{figure}
	\centering 
	\subfloat[Sh-2 168]{
		\fbox{\includegraphics[width=0.48\columnwidth]{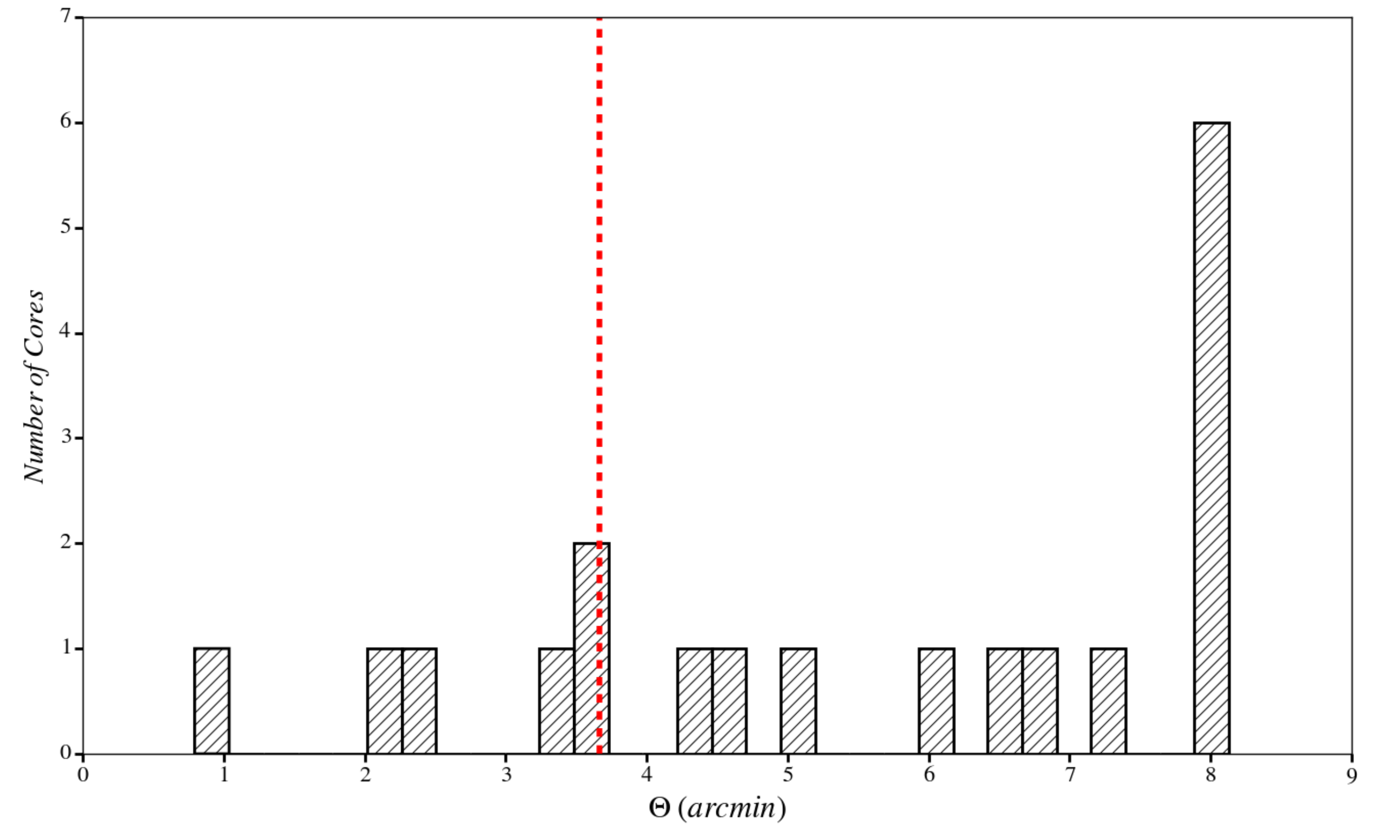}}}
	\subfloat[Sh-2 201]{
		\fbox{\includegraphics[width=0.48\columnwidth]{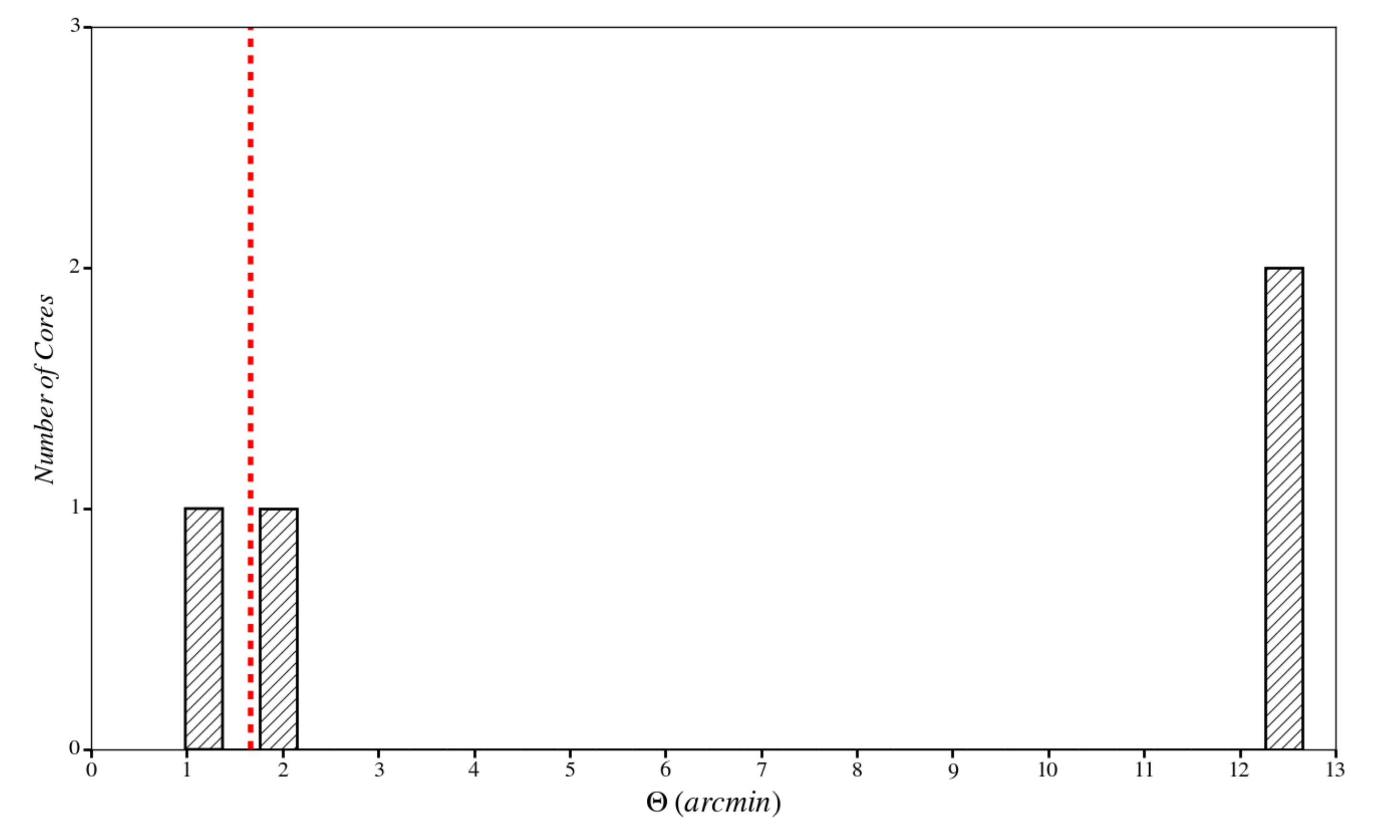}}}
	
	\subfloat[Sh-2 242]{
		\fbox{\includegraphics[width=0.48\columnwidth]{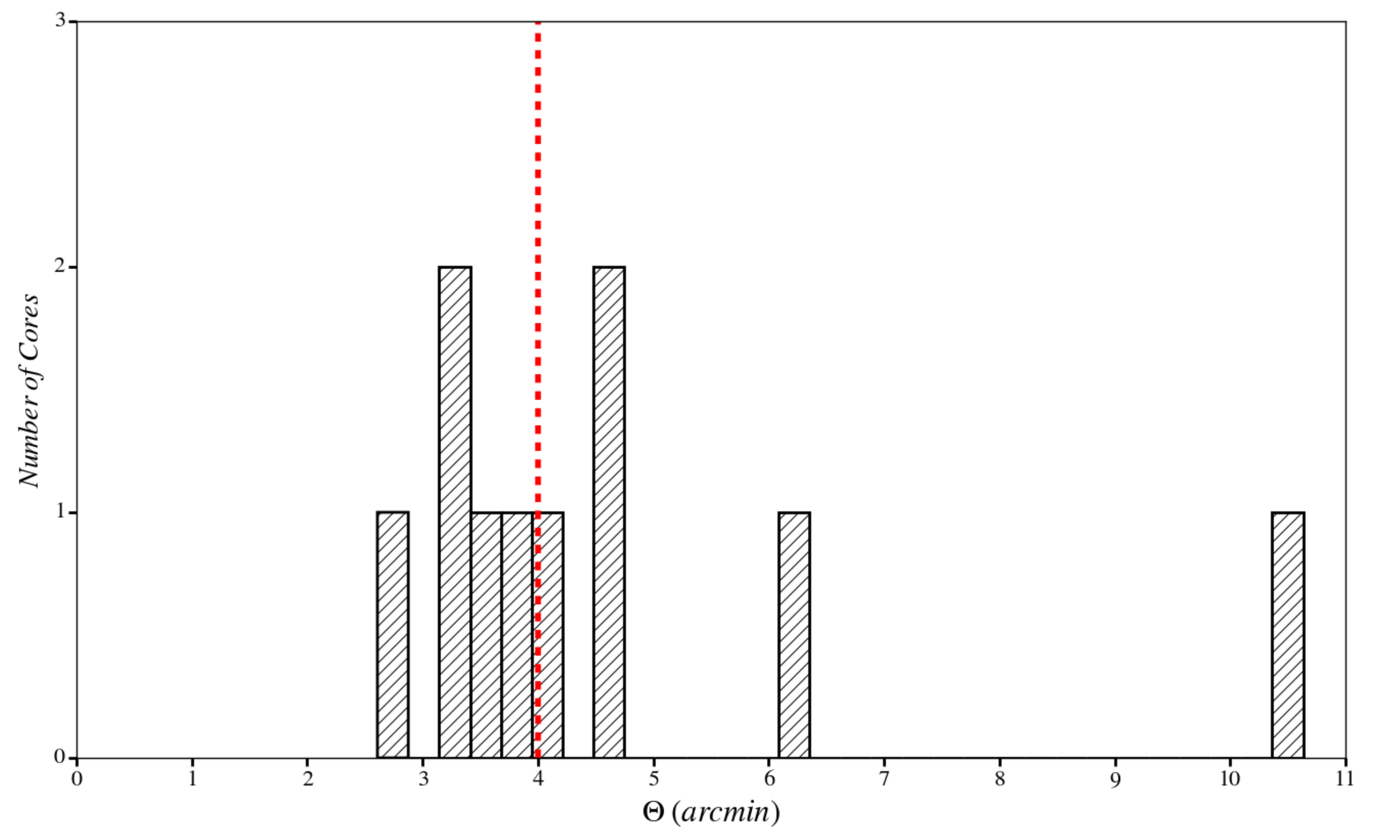}}}
	\subfloat[Sh-2 305]{
		\fbox{\includegraphics[width=0.48\columnwidth]{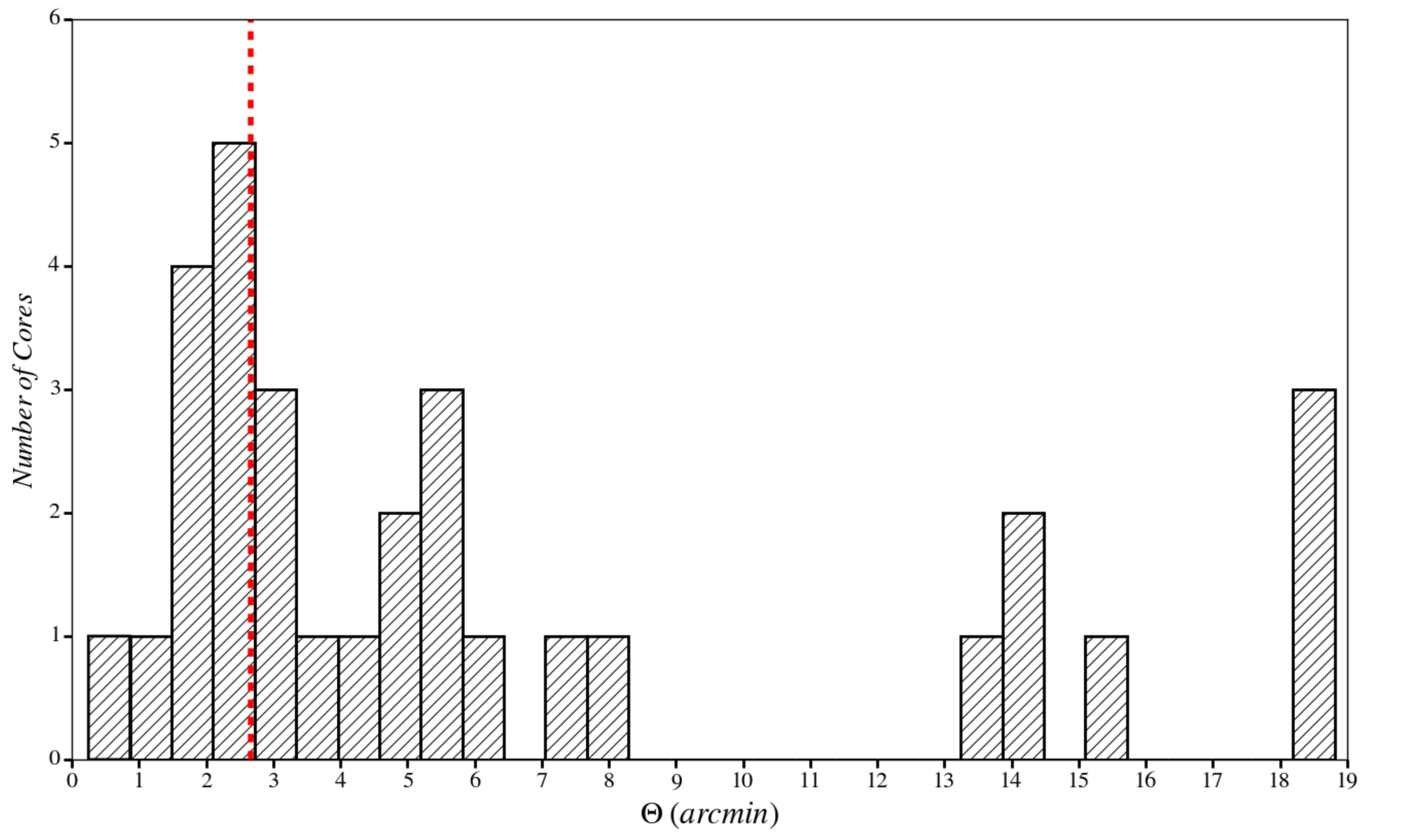}}}
	\caption{The core unscaled radial distribution profiles of the systems presented in Figure \ref{fig:HII_regions}, with 30 equally spaced bins used for each. The dashed red line indicates the boundary of each HII region.}
	\label{fig:HII_region_radial_distributions}	
\end{figure}

As discussed earlier, the systems analyzed in this paper contained a large range of numbers of cores, with only 19\% of the sample containing more than 28 cores. Even the most populated of these systems do not have enough cores for a radial distribution profile to be fitted with any significant degree of certainty.
\\ \\
To further elaborate on this problem, we present the unscaled radial core distribution histogram of the 4 representative cases of Figure \ref{fig:HII_regions} in Figure \ref{fig:HII_region_radial_distributions}. In the first 3 panels we have Sh-2 168 with 19 cores, Sh-2 201 with 4 cores and Sh-2 242 with 10 cores. It is evident that no meaningful radial distribution profile can be fitted due to the small number of cores available. For Sh-2 305, even though it contains 31 cores, the extended condensation along the HII region boundary is marginally distinguishable, while the large span of unpopulated bins at intermediate distances from the HII region give rise to diverging Poisson counting errors, rendering these locations unusable for the radial profile fitting procedure.
\\ \\
To circumvent this issue, we stacked the counts from all the objects in our entire sample, seeking an average radial distribution fit rather than fitting on a case-by-case basis. Furthermore, to achieve a distance-independent result, we scaled the stacked data by the associated HII region radius; the new distance abbreviated simply as \textquotedblleft scaled separation distance" from here onwards. It should be noted that in the case of multiple HII regions in the vicinity of a core, the associated HII region was selected to be the one that shared the smallest scaled separation distance with the core of interest.
\\ \\
A cutoff was placed at $\Theta_{SCALED} = 12$ to segregate cores that were less likely to be associated with their nearest HII region. This cutoff corresponds to an angular separation distance anywhere between $6'$ and $150'$ (or $5.2 \ pc$ and $313 \ pc$), with the most likely being $25'$ (or $33 \ pc$), although the specific value strictly depended on the radius of the HII region at hand. These distant cores (18 in total) were not considered in further data analysis. Furthermore, the clouds that ended up with no cores associated to an HII region because of this segregation (9 in total) were also not considered in further data analysis.
\begin{figure}[!htb]
	\centering
	\setlength{\fboxsep}{0pt}
	\setlength{\fboxrule}{1pt}
	\fbox{\includegraphics[width=\linewidth]{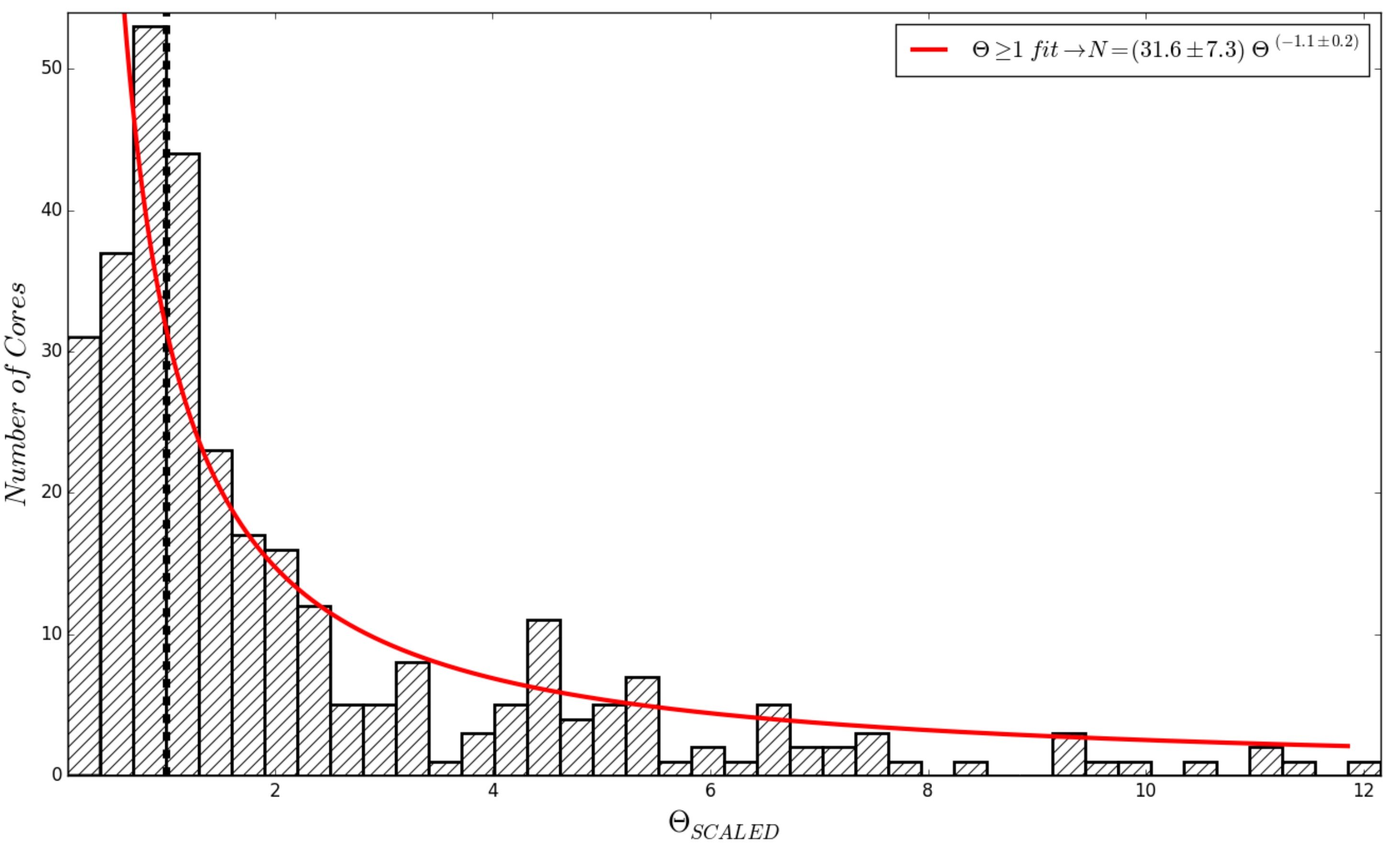}}
	\fbox{\includegraphics[width=\linewidth]{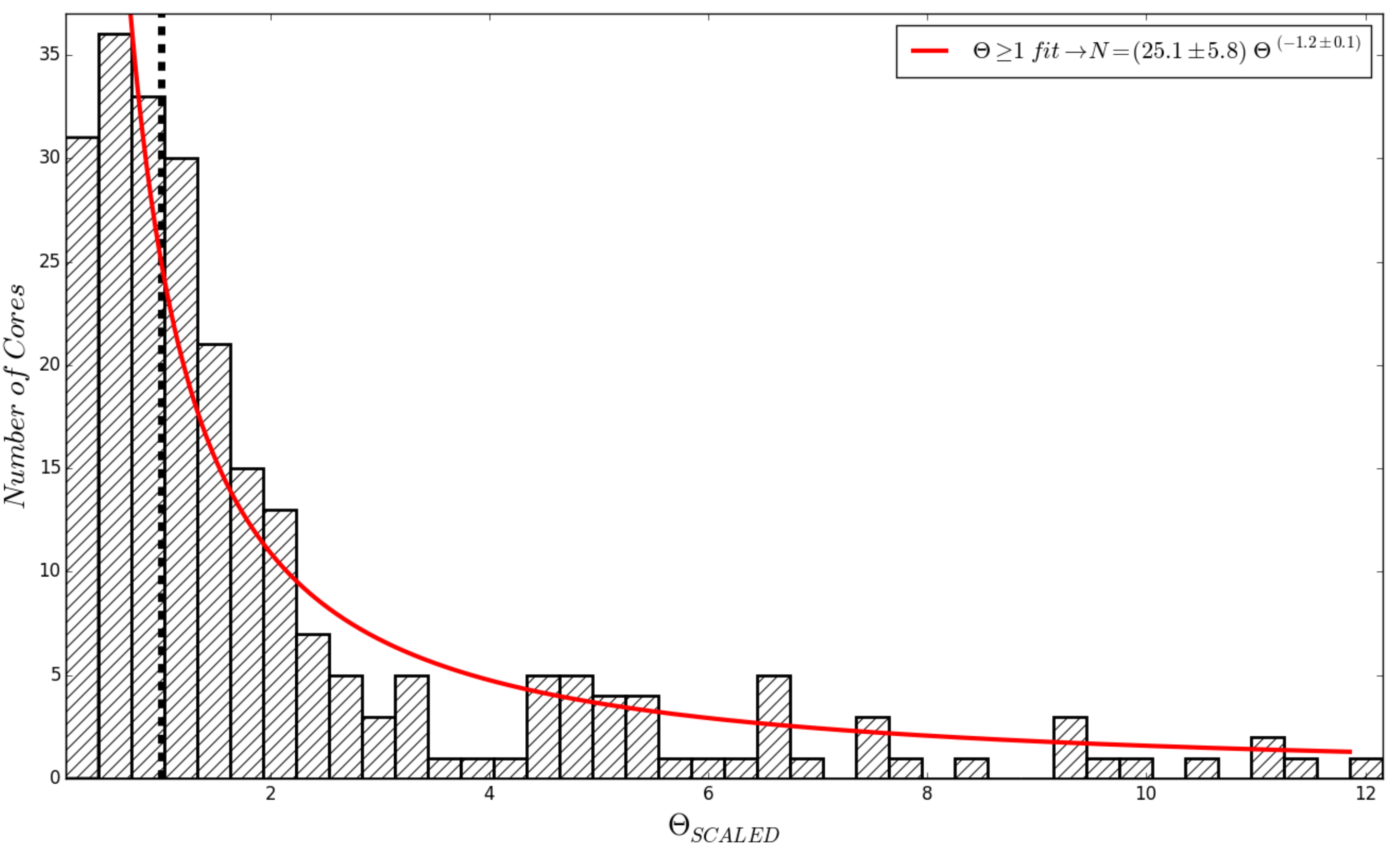}}
	\caption{Two Histograms of core-to-HII Region, center-to-center, separation distances, scaled against the radius of each core's associated HII region. The power-law of best-fit is displayed in red while the black-dashed line indicates the boundary of our HII regions. The top histogram includes the entire core sample, while the bottom histogram excludes the cores from the two \textquotedblleft shell-like" HII regions Sh-2 104 and Sh-2 305.}
	\label{fig:scaled_distance_histogram}
\end{figure}
\\ \\
A scaled separation distance histogram was made for the cores that were within $\Theta_{SCALED} \leq 12$ and is presented in the top plot of Figure \ref{fig:scaled_distance_histogram}. Inspection of this plot reveals that there are many cores far beyond the boundaries of the associated HII regions. The cores comprising these outer bins have been fitted with a power law of the form $N = c_{0} \Theta^{n}$. Different binning options were tested and all cores existing at bins $\Theta_{SCALED} \geq 1$ were used for the fit; the reason for this choice being two-fold. First, this functional form diverges for small $\Theta_{SCALED}$ values, something that is unphysical. Also, the HII regions have almost certainly affected the counts of cores at values $\Theta_{SCALED} \leq 1$.
\\ \\
Overall, the binning option that best represented the data, giving the lowest uncertainty in the fit parameters, was 40 equally-spaced bins. This choice yielded a fit of $N = (31.6 \pm 7.3) \ \Theta_{SCALED}^{(-1.1 \pm 0.2)}$ cores per 12/40 binsize in $\Theta_{SCALED}$, consistent with a volume number density power-law index of -3. This result is robust to variations of the binning used and the minimum $\Theta_{SCALED}$ bin to be included in the fit. The other tested fits were within the error limits quoted with power-law indices $n$ that varied between -1.3 and -0.8. Note that bins with only 1 counted core did not contribute at all to the fitting procedure, as their Poisson counting uncertainty is $\pm 100\%$, which when translated to a log-log scale yields a value of $0^{+0.333}_{-\infty}$. Integration of this fit suggests that out of the 315 identified cores, $70 \pm 3$ (22\%) cores should lie between $1 \leq \Theta_{SCALED} \leq 2$, while the actual count was 90 (29\%), which is significantly more than the expected amount. Fitting only the region $\Theta_{SCALED} \ge 2$ gives a lower curve and the excess number of cores increases substantially but the best fit curve in this case, while still a similar power law (i.e index $\approx -1$), is not so well constrained.
\\ \\
In order to estimate the level of background cores, a detailed investigation has been carried out by Bobotsis (2018), where a toy-model of the form $N = N_{0} \pi \Theta^{2}$ was fitted between $20 \leq \Theta_{SCALED} \leq 25$ as cores identified at/beyond this scaled distance are expected to be background/foreground cores of our sample. The best-fit values for $N_{0}$ were $4 \times 10^{-4}$ and $1.5 \times 10^{-3}$ depending on which of the considered bins were assigned a higher weight. This value of $N_{0}$ translated to a total of 2 to 6 cores expected to be part of the foreground/background in the entire core sample, and a probability of 0.2\% to 0.6\% for encountering a background/foreground core within 20$'$ from the center of any HII region from this sample, two results that rendered the issue of background/foreground contamination insignificant.
\\ \\
Even though we did not fit for $\Theta_{SCALED} \le 1$, we can see that the behavior there is very different than $\Theta_{SCALED} \geq 1$ as expected due to the different environment at hand. Specifically, a dramatic decrease in the number of cores can be seen interior to $\Theta_{SCALED} \le 1$, while an excess is seen near $\Theta_{SCALED} \approx 1$.
\\ \\
To further establish the significance of the observed core excess, the cores contributed from the shell-like HII regions Sh-2 104 and Sh-2 305, which populated mostly $\Theta_{SCALED} \leq 2$, were all removed from the core counts and the resulting histogram is presented at the bottom plot of Figure \ref{fig:scaled_distance_histogram}. A separate power-law fit was made for this histogram for the cores lying outside their associated HII region ($\Theta_{SCALED} \geq 1$). The resulting fit was $N = (25.1 \pm 5.8) \ \Theta_{SCALED}^{(-1.2 \pm 0.1)}$ with power-law indices $n$ that varied between -1.3 and -0.4; a result less robust than when considering the full sample. Integration of this fit made for the cores at $\Theta_{SCALED} \ge 1$ suggests that $53 \pm 3$ (17\%) cores should lie between $1 \leq \Theta_{SCALED} \leq 2$, while the actual count was 75 (31\%), which is greater than the expected amount by a larger and more significant amount than in the complete data set calculation above. It is evident that the observation of a large excess in the number of cores near the boundary of the HII regions ($\Theta_{SCALED} \approx 1$) was unaffected by this experiment, suggesting that the two shell-like HII regions do not introduce any significant bias in the interpretation of the earlier result.
\subsection{Cloud and Core Temperatures}

\begin{figure}
	\centering 
	\subfloat{
		\fbox{\includegraphics[width=0.48\columnwidth]{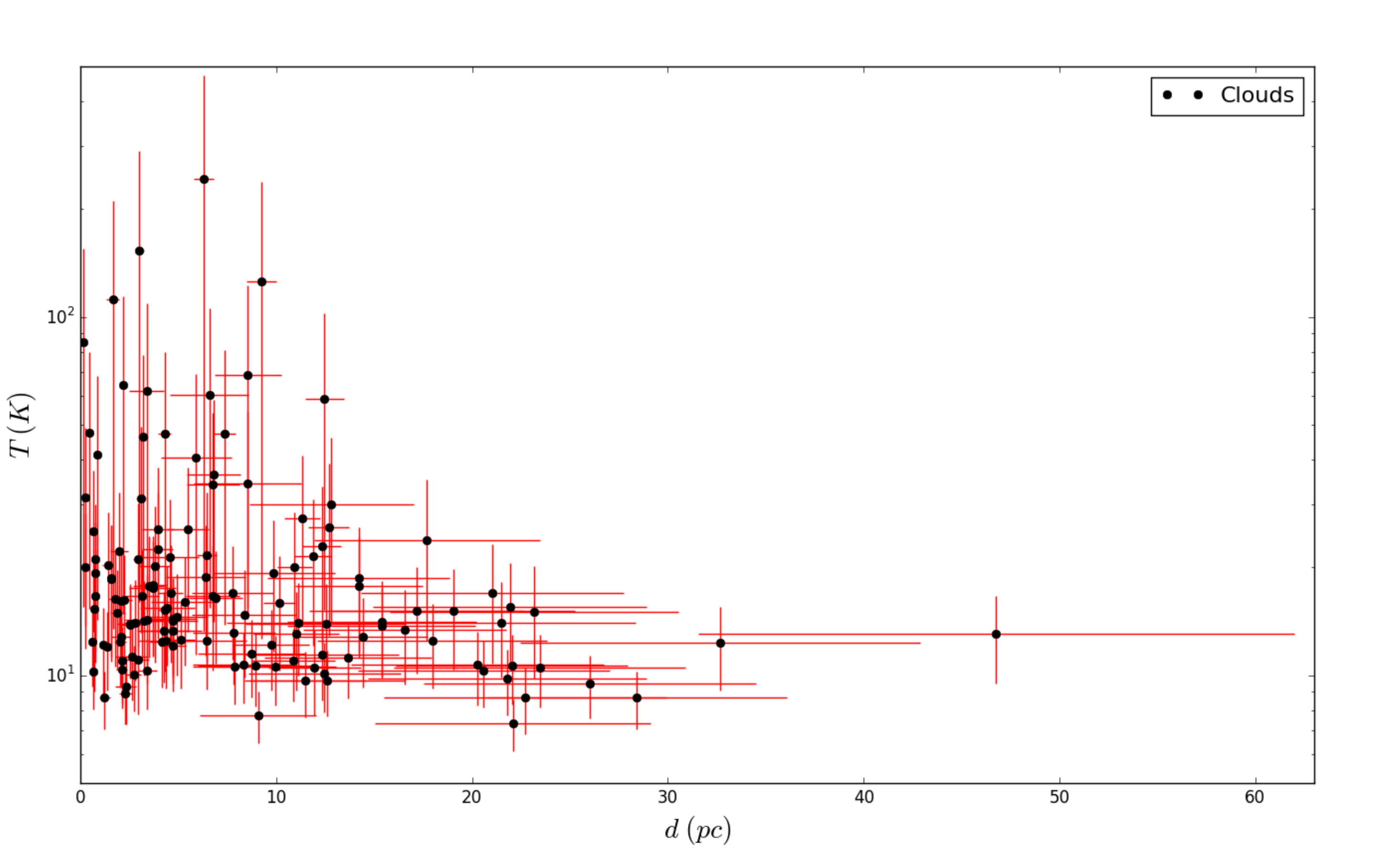}}
	}
	\subfloat{
		\fbox{\includegraphics[width=0.48\columnwidth]{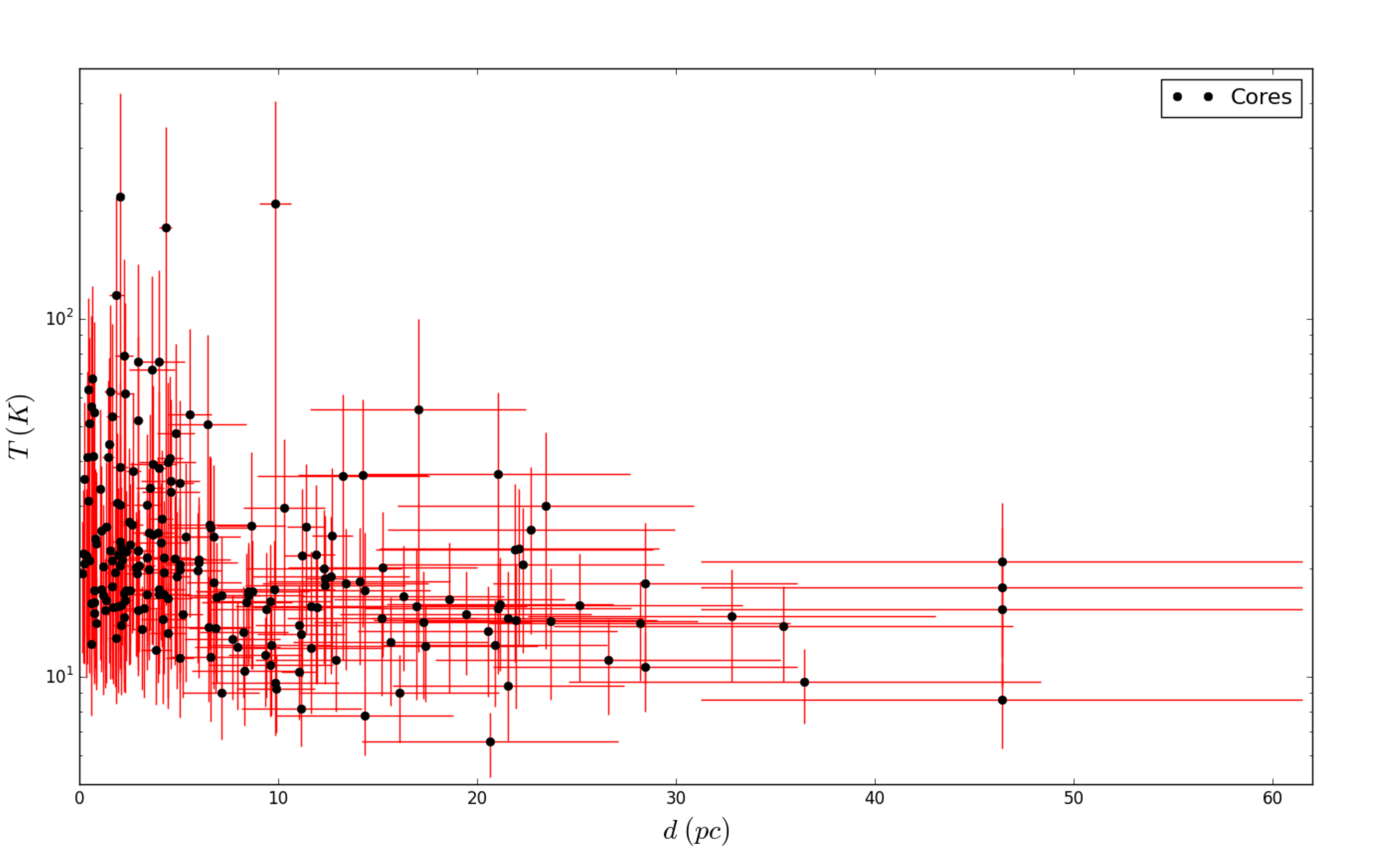}}
	}
	
	\subfloat{
		\fbox{\includegraphics[width=0.48\columnwidth]{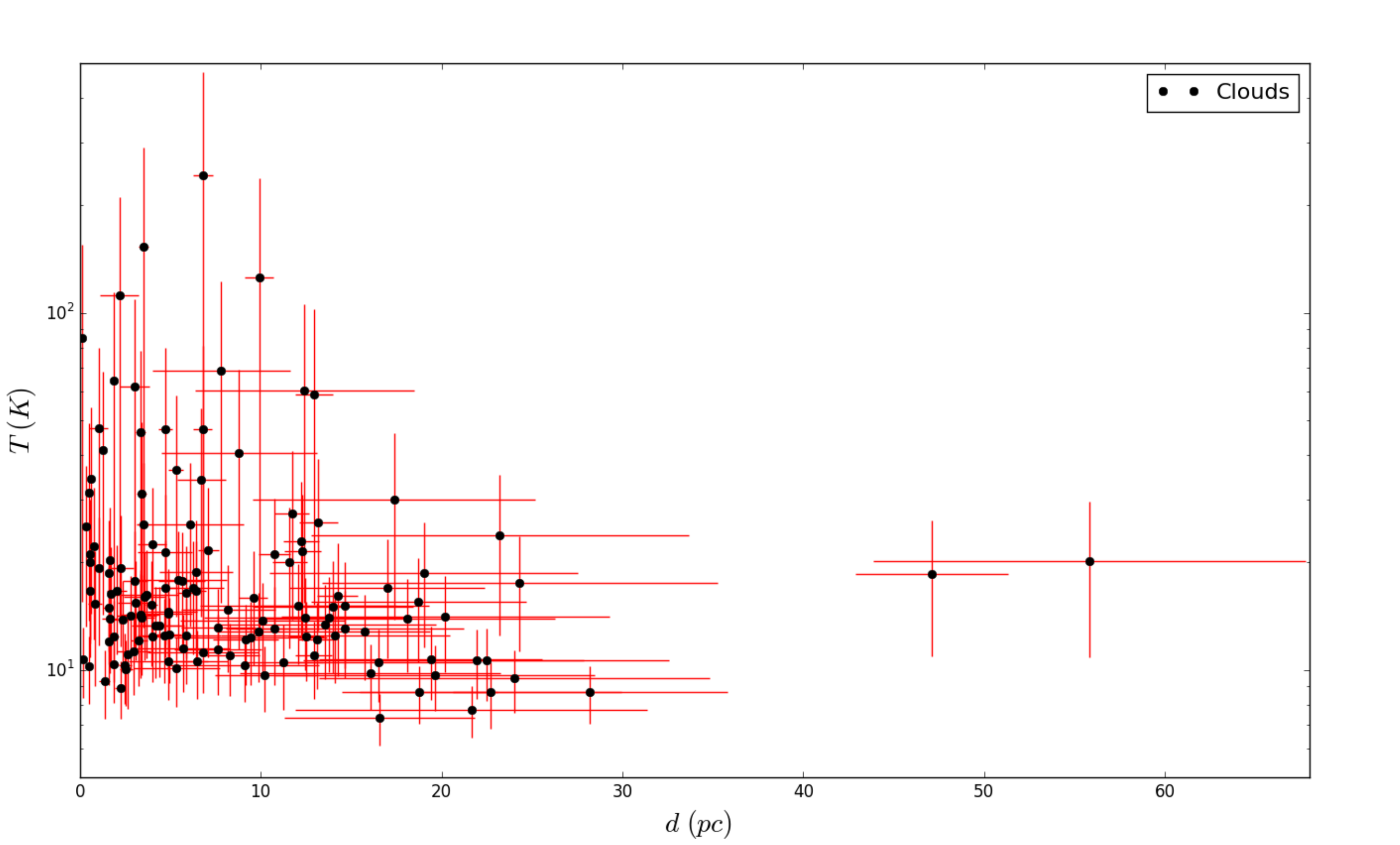}}
	}
	\subfloat{
		\fbox{\includegraphics[width=0.48\columnwidth]{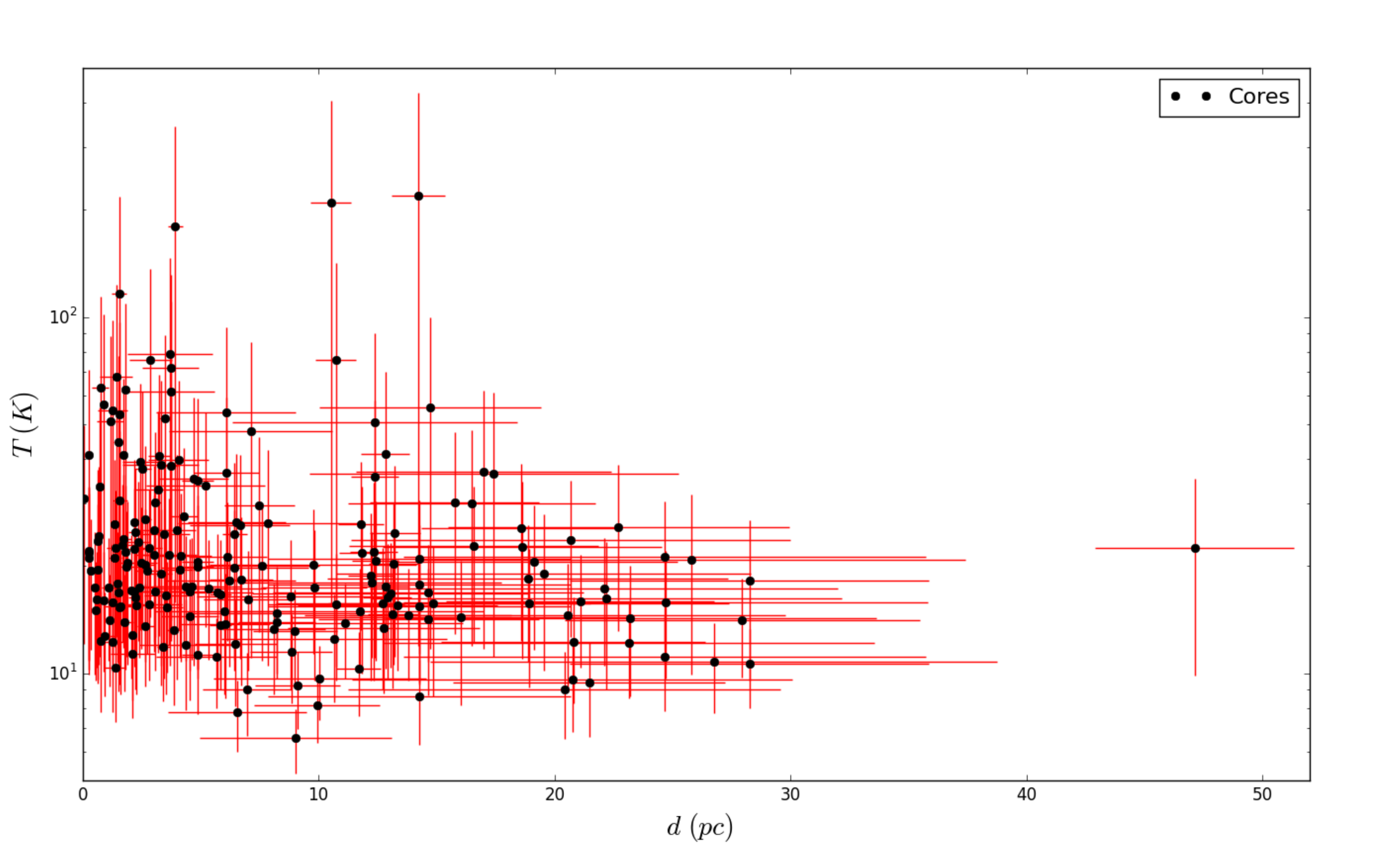}}
	}

	\caption{Average temperature against physical separation distance. In order of appearance, the HII Region - Cloud (Top-Left), HII Region - Core (Top-Right), nearest OB star - Cloud (Bottom-Left) and nearest OB star - Core (Bottom-Right) comparisons are displayed.}
	\label{fig:heating}
\end{figure}
\begin{figure}
	\centering 
	\subfloat{
		\fbox{\includegraphics[width=0.47\columnwidth]{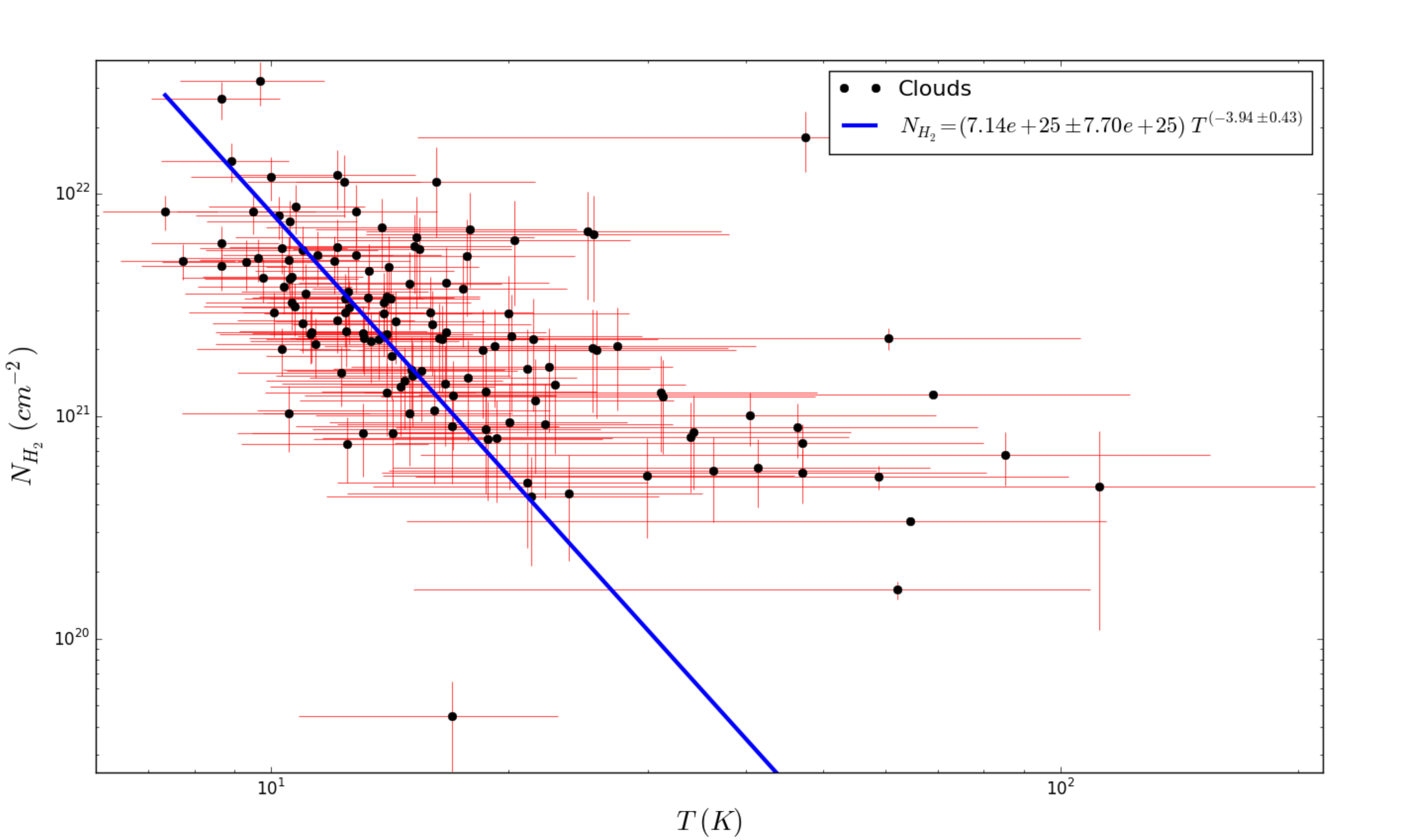}}
	}
	\subfloat{
		\fbox{\includegraphics[width=0.47\columnwidth]{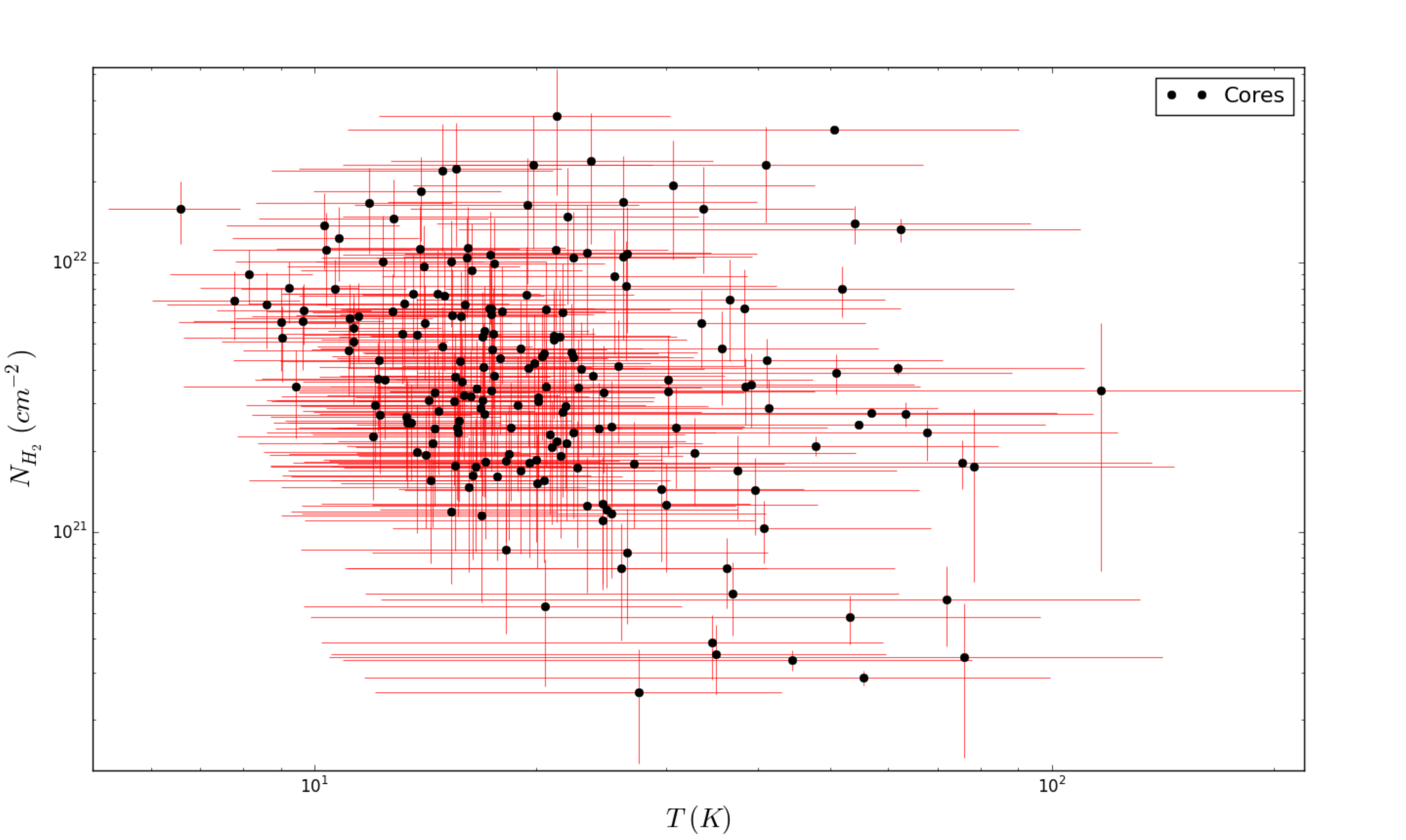}}
	}
	\caption{Average temperature against average $H_{2}$ column density for clouds (Left) and cores (Right).}
	\label{fig:internal_heating}
\end{figure}
\begin{figure}[!htb]
	\centering
	\setlength{\fboxsep}{0pt}
	\setlength{\fboxrule}{1pt}
	\fbox{\includegraphics[width=\linewidth]{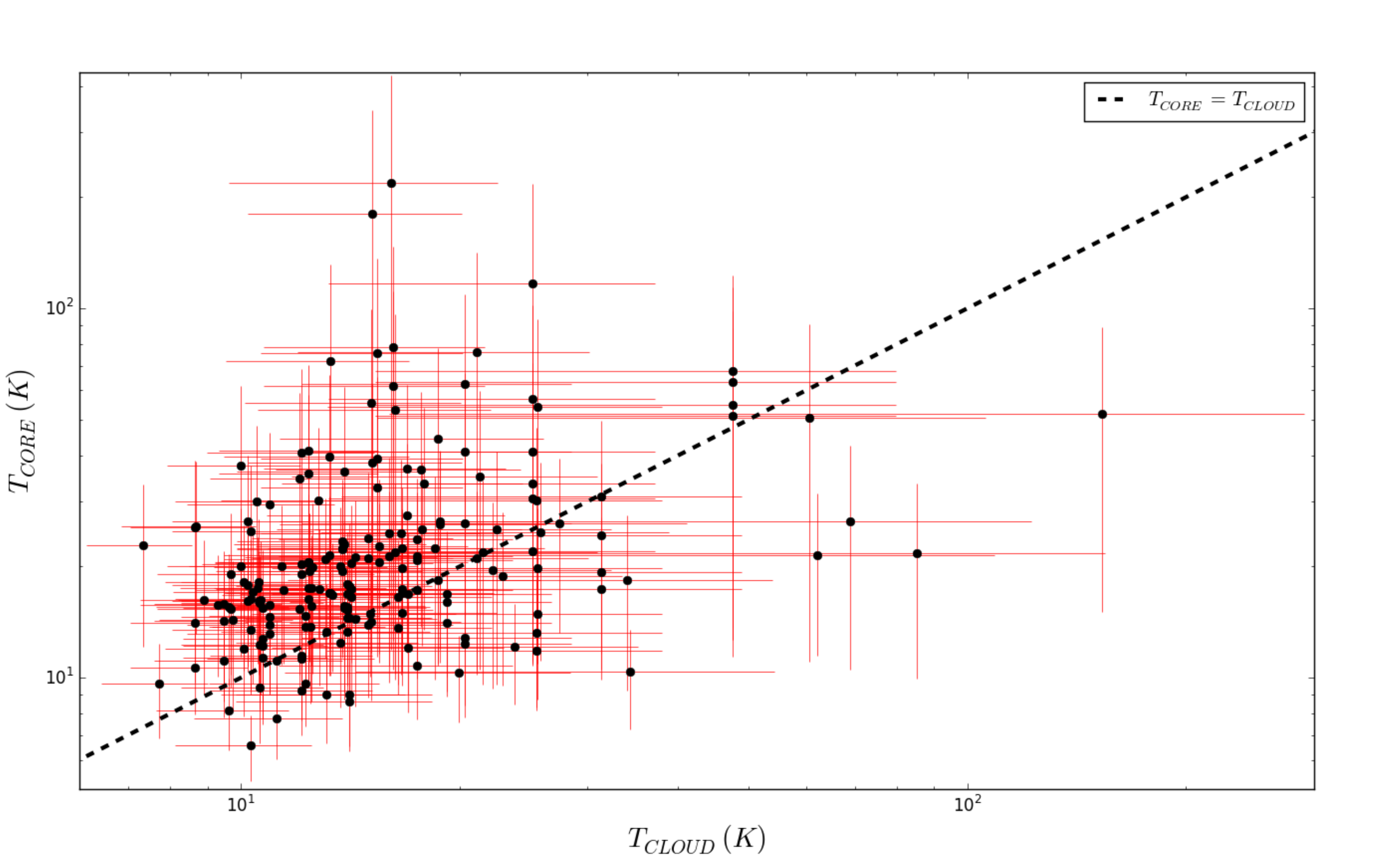}}
	\caption{Average cloud temperature against average embedded core temperature. A black, dashed line is used to display cloud-core temperature equivalence.}
	\label{fig:cloud_core_temp}
\end{figure}

To investigate any heating effect taking place either due to the HII region, or the HII region's parent star(s), the average temperature of each core and cloud is compared against the physical distance between them and their associated HII region, as well as their nearest OB star. The results from these comparisons are presented in Figure \ref{fig:heating}. There is no significant heating effect indicated in any of the relationships plotted in this figure. There is a slight trend apparent to the eye for a larger number of high temperature points at physical separation distances $\leq 10 \ pc$, but the statistical significance of this trend is quite low.
\\ \\
However, a comparison between cloud and core average temperature against average $H_{2}$ column density shown in Figure \ref{fig:internal_heating} does show a dependency of high significance for the clouds. The clouds surrounding these dense cores are generally found to be warmer when they have smaller $H_{2}$ column densities. A power-law fit was made for the clouds and the result was $ N_{H_{2}} = (7.14\times 10^{25} \pm 7.70\times 10^{25}) \ T^{(-3.94 \pm 0.43)}$. This is not surprising, since a lower column density corresponds to a lower extinction, which in turn means a greater penetration for incoming photons from nearby stars. This observation is consistent with various cooling models for molecular clouds which suggest a negative power-law dependency to both column and number density (Juvela, Padoan, \& Nordlund 2001). On the other hand, the equivalent comparison for the cores does not show any convincing correlation between average temperature and average $H_{2}$ column density.
\\ \\
We compared the average temperature of all cores to the clouds surrounding them and present the result of this comparison in Figure \ref{fig:cloud_core_temp}. Out of the 315 cores considered, we were able to measure a temperature for 199 (63\%) of these cores and their surrounding cloud. Of these 199 cores, 147 (74\%)  were found to be warmer than their surrounding cloud. No significant correlation between core and cloud average temperature was found.

\pagebreak
\subsection{Star Formation Efficiency}
The dust emission measured by the SCUBA-2 instrument provides a sensitive measure of the total mass of material around the young stars near the HII regions considered in this sample. Consequently, it is possible to use the masses derived from SCUBA-2 measurements to determine the \textquotedblleft Star Formation Efficiency" (SFE) for these systems. The SCUBA-2 measurements provide a separate measure from that of spectral line observations of molecules, such as CO.
\\ \\
The SFE ($\epsilon$) was determined for HII region systems with a complete, or almost-complete mass budget by simply comparing their gaseous and stellar mass budgets in the following manner:
\begin{equation}
\epsilon = 100 \times \bigg( \frac{M_{STAR}}{M_{STAR} + M_{GAS}} \bigg)
\end{equation}
For the gaseous component, the mass of the ionized gas plus the masses of all clouds and cores are summed together. The ionized gas mass is generally less uncertain than the cloud and core masses because the estimates of the latter are derived from dust mass which can sometimes suffer excessively from noisy 450$\mu$m photometry.
\\ \\
For the stellar component, the mass of the massive OB stars is summed separately from that of the low and intermediate-mass stars, which is estimated using a Kroupa (2001) \textquotedblleft Initial Mass Function" (IMF). A maximum stellar mass must be set to use the Kroupa IMF. For each object we have determined the OB stars associated with the HII region. We only calculate the SFE for those objects where we find such OB stars located within or near enough the HII region to be the exciting stars. We use the least massive OB star in each HII region as the upper limit for the Kroupa IMF determination of the mass of the stars with lower masses.
\\ \\
This assumes that all HII region systems have a complete account of their associated, massive OB stars, and consequently the mass contributed from these. Identifying the OB stars associated with each HII region is likely the largest uncertainty contributor in the stellar mass budget. The calculated SFE values are presented in Table \ref{table:SFE}, with detailed descriptions of each system, including image diameters and 450$\mu$m/850$\mu$m noise-per-pixel values.
\begin{table}[!htb]
	\centering
	
	\caption{Table of HII region systems whose SFE was obtainable from our data. Columns in order of appearance indicate (1) System ID (2) Contained HII regions (3) SFE, and (4) a short description of each system justifying assigned uncertainty. Systems in red are of very high uncertainty.}
	
	\begin{footnotesize}
	\setlength\tabcolsep{2pt}

	\begin{tabular}{|c|c|c|c|}
		
	\hline
	\textbf{System} & \textbf{HII Regions} & \textbf{SFE (\%)} & \textbf{Description} \\
	\hline
	G70 & Sh-2 99,100 & $1.02\pm0.3$ & \specialcell[c]{2 interacting HII regions, primary targets, many HM stars \\ many filaments, $1800''$, $[N_{450}, N_{850}] = [5, 1.3]$ mJy/beam} \\
	\hline
	G74 & Sh-2 104 & $7.85\pm1.5$ & \specialcell[c]{1 HII region, primary target, several HM stars, few filaments \\ $1800''$, $[N_{450}, N_{850}] = [5.7, 0.3]$ mJy/beam} \\
	\hline
	G97 & Sh-2 128 & $2.37\pm0.5$ & \specialcell[c]{1 HII region, primary target, 1 HM exciting star, several filaments \\ $1800''$, $[N_{450}, N_{850}] = [5.8, 1.3]$ mJy/beam} \\
	\hline
	G108 & Sh-2 152 & $7.97\pm1$ & \specialcell[c]{1 HII region, legacy survey, 1 HM exciting star,\\ few filaments, $3600''$, $[N_{450}, N_{850}] = [81.2, 3.4]$ mJy/beam} \\
	\hline
	G115 & Sh-2 168 & $29.5^{+2}_{-6}$ & \specialcell[c]{1 HII region, primary target, many HM stars in vicinity, \\ lots of filaments, $1800''$, $[N_{450}, N_{850}] = [4.7, 1.4]$ mJy/beam} \\
	\hline
	G120 & Sh-2 175 & $26.2^{+2}_{-5}$ & \specialcell[c]{1 HII region, primary target, 1 HM exciting star, \\ several filaments, $240''$, $[N_{450}, N_{850}] = [25.6, 1.2]$ mJy/beam} \\
	\hline
	G173 & \specialcell[c]{Sh-2 231,232\\233,235} & $19.2\pm4$ & \specialcell[c]{2 interacting HII regions, legacy survey, several HM stars in vicinity, \\ several filaments, $3600''$, $[N_{450}, N_{850}] = [138,3.4]$ mJy/beam} \\
	\hline
	\color{red} G173B & \color{red} \specialcell[c]{Sh-2 234,237} & \color{red} $72.6^{+2}_{-30}$ & \color{red} \specialcell[c]{2 non-interacting HII regions, legacy survey, lots of HM stars without \\ certain association, few filaments, $3600''$, $[N_{450}, N_{850}] = [187, 3.3]$ mJy/beam} \\
	\hline
	G182 & Sh-2 242 & $9.10\pm2$ & \specialcell[c]{1 HII region, primary target, 1 HM exciting star, \\ several filaments, $1800''$, $[N_{450}, N_{850}] = [8.1, 1,7]$ mJy/beam} \\
	\hline
	G188 & Sh-2 247 & $8.09\pm3$ & \specialcell[c]{1 HII region, legacy survey, 1 HM exciting star, \\ several filaments, $3600''$, $[N_{450}, N_{850}] = [16.5,0.8]$ mJy/beam} \\
	\hline
	G192 & \specialcell[c]{Sh-2 254,255,\\256,257,258} & $6.65^{+4}_{-1}$ & \specialcell[c]{5 interacting HII regions, legacy survey, several HM stars, \\ several filaments, $3600''$, $[N_{450}, N_{850}] = [11.4, 3.0]$ mJy/beam} \\
	\hline
	G192B & Sh-2 255B,259 & $3.41\pm1$ & \specialcell[c]{2 non-interacting HII regions, legacy survey, 1 HM exciting star, \\ few filaments, $3600''$, $[N_{450}, N_{850}] = [11.4,3.0]$ mJy/beam} \\
	\hline
	G210 & Sh-2 283 & $29.7^{+3}_{-8}$ & \specialcell[c]{1 HII region, primary target, few HM stars with uncertain association, \\ several filaments, $240''$, $[N_{450}, N_{850}] = [95.5, 4.5]$ mJy/beam} \\
	\hline
	\color{red} G219 & \color{red} Sh-2 288 & \color{red} $69.1^{+5}_{-35}$ & \color{red} \specialcell[c]{1 HII region, legacy survey, 1 HM star, unknown filamentary amount, \\ $7200''$, $[N_{450}, N_{850}] = [106, 4.6]$ mJy/beam} \\
	\hline
	G221 & BFS 64 & $37.5^{+5}_{-15}$ & \specialcell[c]{1 HII region, primary target, few HM stars with uncertain association, \\ few filaments, $240''$, $[N_{450}, N_{850}] = [141,4.5]$ mJy/beam}  \\
	\hline
	G233 & Sh-2 305 & $3.28\pm0.3$ & \specialcell[c]{1 HII region, primary target, few HM stars, several filaments \\ $1800''$, $[N_{450}, N_{850}] = [5.4, 4.2]$ mJy/beam} \\
	\hline
	\label{table:SFE}
	\end{tabular}
	\end{footnotesize}
\end{table}
\\ \\
Of the 31 HII region systems in our sample, only 16 (52\%) had a sufficiently complete gas and star mass budget for an SFE estimate to be made. Of these, a majority of 9 systems displayed SFE values below 10\%, while 5 systems consisted of large-value outliers, and another 2 systems had very incomplete gas mass budgets, leading to only an upper limit estimate for their SFE (indicated in red in Table \ref{table:SFE}). The first of these 2 systems was G173B which had an SFE value of 72.6\%. This system is comprised of 2 HII regions (Sh-2 234 and Sh-2 237), 4 submillimeter-emitting clumps, of which only 1 had a determined total mass, and an unusually large list of 15 potentially associated OB stars, of which several may not be associated with the 2 HII regions of the system. The second of the two systems was G219 which had an SFE value of 69.1\%. This system is comprised of 1 HII region (Sh-2 288), 1 submillimeter-emitting clump and 1 associated OB star. The mass of the single clump is expected to be very underestimated. This is mostly due to poor atmospheric conditions at the time of observation, but also, due to the short integration time per pixel used in the scan itself, something that would most certainly render any low-mass clumps in the system practically undetectable.

\section{Discussion}
In our analysis, even the 6 most populated HII regions did not have enough cores to produce a statistically significant radial profile fit. However, the entire sample of cores, brought together in a scaled fashion as was done in this work clearly shows an extended core population. Attempts to analyze this population using un-binned statistics have so far been unsuccessful, in part due to the presence of a (small) background of cores affecting the large scales and a divergence in the expected number of cores at the smallest scales (e.g one needs to also account for the sizes of the cores, especially at the center of the radial distribution).
\\ \\
The results of this analysis of the distribution are (1) cores well beyond the HII region are distributed around it such that their number is consistent with a spherical population; (2) there is an excess number of dense cores just outside the HII regions, even when the obvious shell-like objects are removed from the sample; (3) the number of dense cores at small distances is consistent with no dense cores existing inside the HII region; (4) the amount of background core contamination is very insignificant.
\\ \\
Regarding our first result, our number counts are given in number of cores within equally spaced circular rings. At larger distances from the HII region, far beyond the ionized gas boundary, the number of cores decreases approximately as $r^{-1}$; therefore the surface density of the cores decreases as $r^{-2}$. We consider two opposite extremes for the large scale distribution of these dense cores: (i) a spherical distribution against a (ii) filamentary structure.
\\ \\
It is easy to show that filamentary structures of common length, with dense cores uniformly distributed along the filament, is a poor fit for the observed radial distribution found here. If filaments have a common (scaled) length (e.g. $L_{max}$), then the number of dense cores in each ring ($N(\Theta_{scaled})$) would follow a functional form proportional to ($L_{max}^{2}$ -   $\Theta_{scaled}^{2})^{1/2}$, which is very different from the observed number counts, even having the opposite curvature in the plot.  However by using more complex filamentary structures, such as a distribution of the lengths of the filaments and/or a non-uniform distribution of the dense cores along the filaments, it is possible to fit these number counts. However, with this level of freedom to choose parameters one could fit almost any number count distribution.
\\ \\
Alternatively if the dense cores are in a spherical distribution at large distances from the HII region, then the determination of the distribution function is simple: the volume density of the dense cores follows a $r^{-3}$ distribution.  There would likely be significant dynamical differences between these two extremes (i.e filament versus spherical distribution). This suggests that a kinematic investigation of these cores (e.g. radial velocity observations) would be very useful in making progress on this question. Furthermore, the spherical distribution model would appear to contradict models in which massive stars are formed near the edges of GMCs, where an external trigger has been applied to start the star formation process. Kirk et al. (2016) have examined clusters of dense cores in Orion B and found that the most massive of these cores is near the center of each \textquotedblleft dense core cluster". They suggest that mass segregation has already occurred before the first star forms, and that the most massive star will then form in the center of this cluster.  This is consistent with our result, where the massive OB star forming an HII region is at the center of a cluster of dense cores.
\\ \\
Our second result that there are additional cores in a shell around the outer edge of the HII region, seems to be true in general, not only for the few obvious, generally well-studied, shell-like HII regions. This may suggest that collect and collapse models are useful to describe most HII regions. Perhaps the lack of obvious shells around many HII regions is simply the result of lower densities of surrounding material.
\\ \\
The idea that these HII regions have shells of dense cores around and immediately outside the ionized region does not contradict the selection criteria that these HII regions are all very obvious in visible light.  In theory these shells are expected to be far from uniform in density with strong instabilities causing the production of denser regions with these shells.  This is observed.  Many, perhaps most, visible HII regions show patches at a few positions where the emission from recombination is completely absorbed. However, when averaged over much of their surface area, the extinction to these HII regions is typically only a few magnitudes as seen in many studies including from measurements of the Blamer decrement over large apertures (Fich and Silkey, 1991).  Even deeply embedded, presumably younger HII regions show the neutral material in very clumpy structures with spherical shells in 3D (Topchieva et al., 2018,2019).
\\ \\
This is further complicated by the observation that many HII regions that are seen in the visible are on the near side of large molecular clouds and emerging towards the observer, with much less neutral material on the near side than on the far side. However, even these HII regions will still show strong enhancements in the numbers of dense cores along the edges of the visible region when seen in projection on the sky, with smaller numbers, from the far side, seen at smaller projected distances.
\\ \\
Our third result from the core radial distribution is that the number of dense cores within the HII region boundary is small, and not of great surprise. There will be a significant number of cores seen in projection through the HII regions from the large scale distribution and from a shell around the boundary of the HII regions. The number seen from the boundary shell, through or in front of the ionized gas, depends on the thickness of the shell, and the number in this shell is always less that the number seen around the projected edges of the HII region. There is no need to match the number counts for there to be any dense cores within the ionized gas region. This does not mean that such objects (dense neutral cores) will never be seen within HII regions but the numbers strongly suggest that these should be rare. Any dense cores from the initial cluster that find themselves within the HII region will be eventually destroyed by the action of the ionizing star. In our sample of older, mature HII regions one would expect this core destruction process to be quite advanced.
\\ \\
Our result on the temperatures of the cores and their surrounding clouds was unexpected. Structures at the edge of the HII region; perhaps impacted by their expanding shock; or closer to the very luminous OB stars in, or near the HII region, might reasonably be expected to be at higher temperatures than more distant clouds and cores. However, this is not seen in this work, an observation consistent with recent work from Rumble et al. (2014)  where the main B-type exciting star MCW 297 was found to have no noticeable heating effect on any but its nearest clump, with which it shared a small physical separation distance of 0.05 pc.
\\ \\
We also searched for correlations of average temperature with other properties, such as projected stellar heating flux (i.e luminosity of nearby OB stars divided by projected physical distance squared). The only correlation we found was that the clouds were hotter when they had lower column densities. Taken together, these results are consistent with cloud heating being dominated by the diffuse interstellar radiation field and not necessarily by any one nearby star. One caveat to this result is that our sample does not include any HII regions that are heated by the most luminous earlier-type O stars.
\\ \\
The cloud $T$ and $N_{H_{2}}$ correlation also suggests that a \textquotedblleft shielding" effect is in-place, where essentially the outer cloud layer allows progressively less external radiation from reaching the inner core condensations by virtue of its extinction, which in-turn prevents the interior cores from being heated to any significant extent by external radiation.
\\ \\
Our last temperature result was that most cores ($\approx 74\%$) were hotter than the cloud that surrounds them. There are many on-going searches for cold cores being the progenitors of star forming events. However it appears that such cold cores are relatively rare near HII regions. This suggests that the presence of the HII region has caused most of the nearby cores to begin collapse, becoming warmer in the process. An alternative is that the cores are all on similar schedules, beginning to form into stars at the same time, but the most massive core has evolved faster, producing an OB star and consequently an HII region which attracts our attention to investigate that part of the sky.
\\ \\
Our final result involves the calculation of the \textquotedblleft Star Formation Efficiency" (SFE). This required a determination of the mass in stars compared to the total mass of the system (i.e stars + gas). The measurements described here are amongst the few to use the dust emission to measure the gas mass. The dust traces both the atomic and molecular material, an advantage over molecular spectral line studies. However, on larger scales the submillimeter emission is faint and larger uncertainties arise as a result. Identifying all of the stellar mass is also problematic, as the parent stars for some HII regions are not seen, while for some others several candidates exist. Because of these difficulties we were only able to reliably estimate the SFE for a small fraction of our HII region systems.
\\ \\
About half (9) of these HII region systems had very reliable mass budgets and collectively suggested SFE values lower than 10\%. On the other hand a very small portion (2) of these systems had unreasonably high SFE estimates, likely due to the much lower reliability of their gas mass budget as compared to the rest of the sample. Nonetheless, a considerable fraction (5) of these HII region systems with modestly reliable mass budgets suggested larger-than-typical SFE values (19.2, 26.2, 29.5, 29.7 and 37.5 $\%$). It may be important that for these 5 systems most of the identified ionizing stars are B-type, with only 3 consisting of earlier type, more luminous stars (O9V or O9.5V). It should be noted that the lowest SFE estimates were generally made in systems with earlier type stars (O5V, O7V).
\\ \\
In order to validate the use of a Kroupa IMF in the determination of SFE values in the vicinity of ionized gas, we focus on the Sh-2 254 complex, which is the only system investigated sufficiently by other authors to allow for direct comparisons to be made. In our present work, we establish the mass of the ionized gas in the complex to be $\approx$ 127 $M_{\odot}$ using VLA 1.46 GHz data (Bobotsis, 2018). In addition, a lower limit of 3600 $M_{\odot}$ is placed on the total $H_2$ mass by summing the $H_2$ mass present within our identified SCUBA-2 clumps (Bobotsis, 2018). These two values together provide a lower limit of 3727 $M_{\odot}$ to the total gas mass of the complex. A total star mass of $\approx$ 265 $M_{\odot}$ is obtained by extrapolating the Kroupa IMF backwards from the least massive star associated with the region, assuming sample completeness for all mass ranges above that. This results in an upper limit of 6.6\% to the average SFE of the complex.
\\ \\
In Chavarria et al. (2008) the total gas mass of the complex was found to be $\approx$ 6385 $M_{\odot}$ using $^{13}CO$ and $^{12}CO$ data. The total star mass was found by converting total star counts to mass using the median YSO mass (0.5 $M_{\odot}$). The obtained SFE values vary between 4\% and 54\% across different components of the complex with an uncertainty up to a factor of 2. In order to make these SFE estimates comparable to our result, we determine the mass-weighted average of the various SFE values from Chavarria to be $\approx$ 8.2\%. Our value of 6.7\% is almost identical with this value from the Chavarria et al. (2008) dataset and the difference between the two is considerably smaller than the uncertainty estimated for either value. 
\\ \\
Furthermore, in Lim et al. (2015) an extensive photometric study of the Sh-2 254 complex shows evidence for an IMF of slope -1.6 for the mass range $10 \le M/M_{\odot} \le 100$. This is slightly steeper than Kroupa's -1.3 slope for the same mass range. In addition, a lower limit of 169 $M_{\odot}$ is made for the total star mass contained in the complex. This is also in good agreement with our 265 $M_{\odot}$ estimate.
\\ \\
Finally, in Mucciarelli (Mucciarelli, Preibisch, \& Zinnecker 2011), an extended Chandra X-ray survey of the Sh-2 254 complex shows a population of young stars very similar to that expected from extrapolating Kroupa's IMF from the lower mass limit of the completely determined star sample down to star masses of 0.5 $M_{\odot}$. In summary, the use of a Kroupa IMF to determine the total stellar mass gives a result that is consistent with that found by others working in this field and using somewhat different assumptions. However, caution should be exercised regarding its use as many HII region systems such as the Sh-2 254 complex tend to favor low-mass star production (Lim et al 2015).
\\ \\
Regarding our uncertainty for the SFE values presented in table \ref{table:SFE}, the strong, non-linear dependency of our gas mass calculations to the SCUBA-2 450 and 850$\mu$m flux is expected to significantly dominate. Rigorous calculation of this uncertainty is a complicated statistical problem, due to the involvement of non-Gaussian variables as is discussed in Bobotsis (2018). However, we do have a good understanding of this uncertainty level. SFE values prone to high levels of uncertainty arise from systems that:
\begin{itemize}
\item{Were detected in one of the JCMT Legacy Surveys}
\item{Have an incomplete accounting of massive OB stars}
\item{Contain multiple HII regions}
\item{Contain a lot of filamentary gas structure}
\end{itemize}
Systems that are part of a JCMT Legacy Survey are prone to much higher gas mass uncertainties as compared to those that were specifically targeted as part of a project simply due to lower integration times and consequently much higher noise per pixel levels across an image, a natural consequence of being part of wide-field survey image.
\\ \\
An incomplete budget of massive OB stars influences star mass estimates two-fold. Clearly, an unaccounted massive star would yield a significant change in the total star mass budget. In conjunction to this however, missing such information can interfere with the choice of the upper mass limit used in the IMF for determining the mass of intermediate and low-mass stars in the system.
\\ \\
In addition, the interaction between multiple HII region fronts in a particular system clearly influences clump formation and consequently, star formation in their immediate vicinity. Contrary to isolated HII region fronts, these tend to drive SFEs up due to the enhancement of the collect-collapse mechanism. It is merely because of this that such systems should be considered as a category of their own, and their SFEs not be expected to trace those of the isolated cases.
\\ \\
Finally, cases where filamentary gas structure is more prevalent than clump structure are also prone to high systematic uncertainties. This paper identifies the molecular gas mass in dense clumps near HII regions. One might expect that cumulative clump mass significantly outweighs filamentary gas mass. It appears that this holds true for our sample in general.
\\ \\
Efficiencies of few percent are typical for regions where high-mass star formation is taking place. However, there have been a number of studies that have suggested higher SFE values for these types of systems. A notable example is the case of Sh-2 104 and RCW 79, for which SFE estimates were 40\% and 45\% respectively (Zavagno et al. 2005), while our SFE estimate for Sh-2 104 is only 7.9\%. The large discrepancy is largely due to the mass incorporated in the gas component between the works. In Zavagno et al. (2005) only the mass of the most massive clump is incorporated, while in this work the mass of all clumps near the targeted HII regions, as well as the mass of their ionized gas are all incorporated in the gas component, leading to a substantially lower SFE estimate.
\\ \\
Overall then, we have shown the utility of a larger sample in the determination of the effects from HII regions onto nearby condensations. It would be useful in the future  to investigate HII regions with parent OB stars of earlier types (i.e. earlier than O9)  in search of a heating effect on such condensations. Furthermore, the observation of a larger angular scale about each HII region target would allow a closer investigation of the extended core population found in this paper. Finally, if the cluster of dense cores is gravitationally bound, there should be a signature of this in the velocity dispersion profile of the cluster. Specifically, a decrease should be seen far from the center, while the velocity field of the cores inside the shell component will also be different.
\\ \\
On the systematic side of things, further treatment of contamination sources, including line contamination from molecules such as $CH_{3}OH$ and $SO_{2}$, as well as radio-continuum contamination from the HII regions themselves would provide even more reliable photometry. Finally, incorporation of multiple submillimeter wavelengths would allow the construction of an individual $\beta$ fit tailored to each source, as well as the usage of the band couple with the lowest uncertainty when performing flux ratios to determine temperature and subsequent derivative properties, lowering the uncertainty of the obtained properties overall.
\\ \\
Nonetheless, we have measured the properties of the material around a large sample of HII regions, all of them at a late stage in their evolution. Our sample was moderately uniform in most measured properties. However our sample does not contain any of the very large and luminous HII regions that are traditionally used as star formation tracers on galactic scales. The clouds of interstellar material surrounding the HII regions in our sample are not the Giant Molecular Clouds which receive much attention in such studies. The typical masses in this sample are only $\approx 1\%$ of a typical GMC mass. However our sample is probably representative of most HII regions. It remains to be seen how these contribute to overall star formation budgets as compared to the contributions of the small number of very large HII regions associated with the largest GMCs.
\section{Acknowledgments}
The James Clerk Maxwell Telescope has historically been operated by the Joint Astronomy Centre on behalf of the Science and Technology Facilities Council of the United Kingdom, the National Research Council of Canada and the Netherlands Organization for Scientific Research. Additional funds for the construction of SCUBA-2 were provided by the Canada Foundation for Innovation.
\pagebreak

\end{document}